\newif\ifNotes\Notestrue
\newif\ifAnon\Anonfalse
\newif\ifDraft\Draftfalse
\newif\ifCCS\CCStrue
\newif\ifArxiv\Arxivfalse

\documentclass[nonacm,sigconf,screen]{acmart}
\usepackage{appendix}

\usepackage{booktabs} 
\usepackage{xspace}
\usepackage{hyperref}
\usepackage[flushleft]{threeparttable}
\usepackage{pifont}
\usepackage{courier}
\usepackage{graphicx}
\usepackage{array}
\usepackage{soul}
\usepackage{bm}
\usepackage{listings}
\usepackage{float}
\usepackage{subcaption}
\usepackage{oplotsymbl}

\usepackage{enumitem}
\usepackage{calc}
\usepackage{placeins}
\usepackage{tikz}
\usepackage[english]{babel}
\usepackage{blindtext}
\usepackage{marginnote}
\usepackage[nameinlink,capitalise,noabbrev]{cleveref}
\usepackage[ruled,section]{algorithm}
\usepackage{algpseudocode}
\usepackage[framemethod=TikZ]{mdframed}
\crefformat{equation}{#2Equation~#1#3}

\setitemize{noitemsep,topsep=0pt,parsep=0pt,partopsep=0pt,leftmargin=1em}

\definecolor{linkcolor}{rgb}{0.65,0,0}
\definecolor{citecolor}{rgb}{0,0.4,0}
\definecolor{urlcolor}{rgb}{0,0,0.65}

\definecolor{cyan}{rgb}{0.0,1.0,1.0}
\colorlet{lightcyan}{cyan!60}
\colorlet{lightyellow}{yellow!60}
\colorlet{lightgreen}{green!60}

\hypersetup{hyperindex=true,pdfpagemode=UseNone,pdfstartview=FitH,colorlinks=true, linkcolor=linkcolor, urlcolor=urlcolor, citecolor=citecolor}

\newcommand{\swallow}[1]{}
\ifNotes
  \newcommand{\colorcomment}[2]{\leavevmode\unskip\space{\color{#1}[#2]}\xspace}
\else
  \newcommand{\colorcomment}[2]{\leavevmode\unskip\relax}
\fi

\ifArxiv

\else

\fi

\makeatletter
\newcommand\InFloat[2]{\@ifundefined{@captype}{#2}{#1}}
\makeatother

\newcommand{\elmo}{\textsc{Elmo}\xspace}

\newcommand{\elmos}{\textsc{Elmo}*\xspace}
\newcommand{\rosita}{\textsc{Rosita}\xspace}
\newcommand{\horosita}{\textsc{Rosita++}\xspace}
\newcommand{\ttest}{$t$-test\xspace}

\newcommand{\mat}[1]{\bm{#1}}
\newcommand{\present}{\textsc{present}\xspace}
\newcommand{\Present}{\textsc{Present}\xspace}

\newcommand{\parhead}[1]{\vspace{3pt plus 1pt minus 1pt}\par\noindent\textbf{#1}\hspace{.75em plus .5em minus .5em}}

\crefname{figure}{Figure}{Figures}
\crefname{lstlisting}{Listing}{Listings}
\Crefname{lstlisting}{Listing}{Listings}
\Crefname{algocf}{Algorithm}{Algorithms}
\Crefname{challenge}{Challenge}{Challenges}

  \newcommand{\authorlist}[1]{#1}
  \newcommand{\nextauthor}{\relax}
  \newcommand{\myAuthor}[4]{
    \author{#1}
    \affiliation{
    	\institution{#2}
    	\country{#4}
    }
    \email{#3}
  }

\copyrightyear{2021}
\acmYear{2021}
\setcopyright{acmcopyright}\acmConference[CCS '21]{Proceedings of the 2021 ACM
	SIGSAC Conference on Computer and Communications Security}{November 15--19,
	2021}{Virtual Event, Republic of Korea}
\acmBooktitle{Proceedings of the 2021 ACM SIGSAC Conference on Computer and
	Communications Security (CCS '21), November 15--19, 2021, Virtual Event, Republic
	of Korea}
\acmPrice{15.00}
\acmDOI{10.1145/3460120.3485380}
\acmISBN{978-1-4503-8454-4/21/11}

\definecolor{dkgreen}{rgb}{0,0.6,0}
\definecolor{gray}{rgb}{0.5,0.5,0.5}
\definecolor{mauve}{rgb}{0.58,0,0.82}

\lstdefinelanguage[arm]{Assembler}     
{
	keywords={pop,push,ldr,ldrb,str,strb,eors,rors,mov,movs,adds,ands,bics,XXX,PPP,nop,lsls},
	comment=[l]\;
}

\newcommand{\lstdefaultset}{\lstset{
	language=[arm]assembler,
	aboveskip=5mm,
	belowskip=5mm,
	xleftmargin=0mm,
	showstringspaces=false,
	columns=flexible,
	basicstyle={\small\ttfamily},
	numbers=none,
	keepspaces=false,
	numberstyle=\color{gray},
	keywordstyle=\color{blue},
	commentstyle=\color{dkgreen},
	stringstyle=\color{mauve},
	breaklines=true,
	breakatwhitespace=true,
	captionpos=b,
	tabsize=3,
	frame=1
}}

\newcolumntype{P}[1]{>{\centering\arraybackslash}p{#1}} 
\newcolumntype{M}[1]{>{\centering\arraybackslash}m{#1}}

\lstdefaultset

\newcounter{challenge}
\renewcommand{\thechallenge}{C\arabic{challenge}}
\newcommand{\challenge}[2]%
  {%
    \bigskip%
    \begin{mdframed}[roundcorner=6pt, backgroundcolor=black!10]%
      \refstepcounter{challenge}\label{#1}\noindent\textbf{Challenge \thechallenge: }#2%
    \end{mdframed}%
  }

\title{\horosita: Automatic Higher-Order Leakage Elimination from Cryptographic Code}

\begin{document}
	
\fancyhead{}
	
\authorlist{
  \myAuthor{Madura A. Shelton}{University of Adelaide}{madura.shelton@adelaide.edu.au}{Australia}
    \nextauthor
  \myAuthor{{\L}ukasz Chmielewski}{Radboud University and Riscure}{lukasz@cs.ru.nl}{The Netherlands}
    \nextauthor
  \myAuthor{Niels Samwel}{Radboud University}{nsamwel@cs.ru.nl}{The Netherlands}
    \nextauthor
  \myAuthor{Markus Wagner}{University of Adelaide}{markus.wagner@adelaide.edu.au}{Australia}
    \nextauthor
  \myAuthor{Lejla Batina}{Radboud University}{lejla@cs.ru.nl}{The Netherlands}
    \nextauthor
  \myAuthor{Yuval Yarom}{University of Adelaide}{yval@cs.adelaide.edu.au}{Australia}
}

\renewcommand{\shortauthors}{M. A. Shelton, \L. Chmielewski, N. Samwel, M. Wagner, L. Batina, 
  and Y. Yarom}

\begin{abstract}
  Side-channel attacks are a major threat to the security of cryptographic implementations,
  particularly for small devices that are under the physical control of the adversary.
  While several strategies for protecting against side-channel attacks exist,  
  these often fail in practice due to unintended interactions between values deep within the CPU.
  To detect and protect from side-channel attacks, several automated tools have recently been proposed; 
  one of their common limitations is that they only support first-order leakage.

  In this work, we present \horosita, the first automated tool for detecting and eliminating higher-order 
  leakage from cryptographic implementations. 
  \horosita proposes statistical and software-based tools to allow 
  high-performance higher-order leakage detection.
  It then uses the code rewrite engine of \rosita (Shelton et~al.\ NDSS 2021) 
  to eliminate detected leakage.
  For the sake of practicality we evaluate \horosita against second and third 
  order leakage, but our framework is not restricted to only these orders. 

  We evaluate \horosita against second-order leakage with three-share 
  implementations of two ciphers, \present and Xoodoo, and with the 
  second-order Boolean-to-arithmetic masking, a core building block of masked 
  implementations of many cryptographic primitives, including SHA-2, ChaCha and Blake. 
  We show effective second-order leakage elimination at a performance cost of 
  36\% for Xoodoo, 189\% for \present, and 29\% for the 
  Boolean-to-arithmetic masking.
  For third-order analysis, we evaluate \horosita 
  against the third-order leakage using a four-share synthetic example 
  that corresponds to typical four-share processing. \horosita correctly 
  identified this leakage and applied code fixes. 
\end{abstract}

\maketitle

\newcommand{\nbcitep}[1]{~\citep{#1}}
\newcommand{\nbcitet}[1]{~\citet{#1}}

\newcommand{\highlight}[1]{%
	\colorbox{red!50}{$\displaystyle#1$}}

\section{Introduction}
Cryptography is one of the main tools used to protect data, both in transit and at rest.
With the increased proliferation of small computing devices into every aspect of modern life, secure cryptography is more important than ever.
Traditionally, the security of cryptographic primitives was evaluated in terms of their mathematical function.
However, in 1996 \citet{Kocher96} demonstrated that the computation of a cryptographic primitive can interact with the environment in which it is computed.
Such \emph{side channels} can leak information about the internal state of the computation, leading to a potential collapse of the security of the implementation.

Since then, significant effort has been invested in researching side-channel attacks.
On the offensive side, attacks have been demonstrated against various types of primitives, including 
symmetric ciphers~\cite{Bernstein2005,MoradiMP13},
public-key systems~\cite{MessergesDS99,GenkinPPTY16}
post-quantum cryptography~\cite{AysuTTGO18},
and non-cryptographic implementations~\cite{BatinaBJP19,YanFT20,abs-2006-15007,ShustermanKHMMO19}.
These attacks exploit various side channels, such as
power consumption~\cite{Kocher1999}
electromagnetic emanations~\cite{QuisquaterS01,GMO01},
microarchitectural state~\cite{GeYCH18,abs-2103-14244, BZB+05},
and even acoustic and photonic emissions~\cite{GenkinST14,KN+13}.
On the defensive side, proposals range over
hardware designs that reduce emissions~\cite{ChenZ06},
software solutions that ensure secret-independent execution~\cite{GeYCH18},
adding noise to hide the signal~\cite{MoradiM13},
and information masking techniques~\cite{ChariJRR99,IshaiSW03,DBLP:conf/icics/NikovaRR06}.

Masking techniques, in particular, are considered promising because they provide a theoretical 
basis that guarantees protection.
In a nutshell, these operate by splitting secrets into multiple \emph{shares}, such that to recover a secret, an attacker needs to observe all of the shares that comprise the secret.
For example, in order-$d$ Boolean masking, a secret $v$ is split into $d+1$ shares, $v_0,\ldots,v_d$ such that for $1\leq i\leq d$, $v_i$ is chosen uniformly at random, and $v_0$ is selected such that $v_0 = v \oplus v_1 \oplus v_2 \oplus \cdots \oplus v_d$.
Such schemes are considered safe because attackers are limited in the number of observations they can made on the internal state of the implementations.
Thus a $d$-order secure implementation which consists of $d+1$ shares, is 
secure against an attacker that can observe up to $d$ internal 
values~\cite{IshaiSW03}.

Despite the theoretical security, masked  
implementations
often fail to provide the promised 
resistance in practice. A main cause for this failure is unintended interactions between values 
processed by the hardware, which allow an attacker to observe multiple shares with a single 
observation~\cite{GaoMPO20,Papagiannopoulos17,BalaschGGRS14}.
Thus, to protect against unintended interactions, designers need to first implement the cryptographic primitives aiming for best protection and then go through several rounds of evaluation.
In each such round, the implementation is evaluated for the presence of leakage 
and then tweaked to eliminate observed leakage.
The process usually repeats until no evidence of leakage is 
observed.
This experimental 
process is expensive because it requires significant expertise, both in the 
design of cryptographic primitives, and in setting up and 
performing analysis of hardware 
measurements.

To reduce the effort required for producing side-channel resistant implementations, a designer may elect to use a leakage emulator~\cite{BuhanBYS21,Papagiannopoulos17,Veshchikov14,McCannOW17} instead of evaluating the hardware.
A recent proposal goes one step further and suggests \rosita, a tool that combines a leakage emulator with software manipulation techniques, providing automatic elimination of side-channel leakage~\cite{SSB19}.
However, one limitation of \rosita is that it only provides first-order security and cannot protect against higher-order attacks.
Thus, in this paper we ask the following question:

\smallskip

\begin{center}
  \emph{Can we automatically detect and correct higher-order side-channel\\ leakage from implementations protected with masking?}
\
\end{center}

\subsection{Our Contribution}
In this work we present \horosita,
an extension to \rosita~\cite{SSB19} that 
performs higher-order leakage detection and mitigation.
At its core, \horosita extends the leakage detection and root-cause analysis 
capabilities of \rosita to support high-order analysis.
It then uses the \rosita code rewrite engine to modify the evaluated implementation and eliminate leakage.
While \horosita can analyse and fix code at any order,
in this work we concentrate on second- and third-order leakage.
We do not investigate orders higher than three for the sake of practicality. The 
complexity of third-order side-channel analysis is significant and the analysis requires tens 
of millions of traces (i.e.\ number of side-channel measurements).
We expect that fourth-order analysis would require at least hundreds of 
millions of traces (i.e.\ months of trace acquisition with a similar setup to 
ours), making such analyses impractical in many scenarios. 

Implementing \horosita is far from straightforward.
The main appeal of high-order secure implementations is that high-order analysis is significantly more complex then first-order analysis.
In particular, we identify three main challenges: the impact of the quadratic 
(for second order) and cubic (for third order) increases in trace lengths on 
the statistical tools used for the analysis, the explosion in the amount of 
data that needs be processed both due to the increase in trace length and 
because of the required increase in the number of required traces, and the 
complexities involved in performing multivariate root-cause analyses. 

To address these challenges, we develop statistical 
software tools that allow robust and efficient high-order leakage analysis.
Our software tools can combine and analyse millions of traces each with 
thousands of sample points and perform efficient bivariate and trivariate 
analysis on the combined traces.
We believe that these tools are of independent value for the side-channel 
community and could be used for high-order analysis in a wide-range of cases.

We assess the second-order effectiveness of \horosita with three cryptographic 
primitives, which represent different points in the design space of symmetric 
cryptography.
\Present~\cite{SasdrichB018} is a popular lightweight block cipher with a traditional substitution-permutation network design.
We extend the two-share \present implementation of \citet{SasdrichB018} to support three shares.
In contrast, Xoodoo~\cite{DaemenHAK18C, DaemenHAK18} 
is a modern cryptographic primitive that underlies multiple 
higher-level primitives~\cite{DaemenHAK18}.
We implement a three-share version of Xoodoo, building on the non-linear
$\chi$ layer from Keccak. 
Finally, we evaluate Boolean-to-arithmetic masking~\cite{GoubinCHES2001} which 
is a cryptographic building block that converts a Boolean mask to an arithmetic 
mask, and is often required in implementing side-channel resistant instances of 
cryptographic algorithms 
that mix Boolean and arithmetic operations, e.g.,
\mbox{SHA-2}~\cite{FIPS1804},
ChaCha~\cite{Bernstein2008}, Blake~\cite{Aumasson2009SHA3PB}, 
Skein~\cite{Sklein2010}, IDEA~\cite{Lai90}, and RC6~\cite{Rivest1998}. 
We implemented the second-order Boolean-to-arithmetic masking of \citet{HutterJCEN2019}. 

We show that \horosita removes all leakage detected in the real experiment up to 2 million traces in Xoodoo and Boolean-to-arithmetic masking. 
For \present all but one leakage point were removed. 
Further, we find that \horosita only requires to emulate 500,000 traces to achieve the same level of protection as achieved 
by analysing 
2 million side-channel traces from physical hardware.
\horosita is available as an open-source project at \url{https://github.com/0xADE1A1DE/Rositaplusplus}.

In summary, in this work we make the following contributions:
\begin{itemize}[nosep,left=0pt]
  \item We explore automated tools for automatic second and third order side-channel detection and protection. (\cref{s:challenges}.)
  \item We develop statistical and software tools for addressing the challenges. (\crefrange{s:statconf}{s:rootcause}.)
  \item We build \horosita, the first tool to automatically detect and remove unintended 
  high-order leakage,
  evaluate it on three 
  cryptographic primitives and demonstrate its efficiency. (\cref{s:evaluation}.)
  \item We made \horosita and the associated tools available as open source.
\end{itemize}

\subsection{Organisation of this paper}
\cref{sec:background} introduces the necessary background on side-channel 
attacks, masking, univariate and multivariate side-channel leakage assessment methods, 
leakage emulators and automatic countermeasures, and statistical tools that we 
use in this work.
In \cref{sec:horosita}, we describe the design for \horosita and in particular how we extend \rosita to higher orders and what the challenges we face. 
We also describe multivariate root-cause analysis and how Rosita improves the code security by code rewrites. 
Subsequently, in \cref{s:evaluation}, we present the results of our evaluation, including both the emulation results and the complimentary side-channel measurement evaluation. 
Finally, in \cref{sec:conclusions} we conclude the paper.

\section{Background}\label{sec:background}

\subsection{Side-Channel Attacks}

Traditional cryptanalysis attacks 
aim to extract secrets from cryptographic algorithms 
by focusing on the mathematical aspects of such algorithms. 
Side-channel attacks, in contrast,
focus on obtaining internal values processed by the algorithm, which are not expected to become public. 
This information is gained by exposing intermediate values of an algorithm through the 
process of collection and analysis of measurements of physical phenomena. Such phenomena include  
timing, power consumption, acoustics, electro-magnetic emanations or properties such as various internal 
states of CPU components. 

\looseness=-1
In 1996, \citet{Kocher96} was the first to publish an exploit of side-channel leakage to recover 
secret  
information that was processed by a cryptographic algorithm. The algorithm in question was 
implemented with high performance in mind, and therefore ran in non-constant time; this allowed the 
timing differences for different inputs to be exploited.
Subsequently, \citet{Kocher1999} used side-channel information from power consumption to 
recover secret information in a new type of attack called Differential Power Analysis (DPA).
In DPA, an attacker calculates a differential trace by finding the difference between averages 
of measured traces of a certain bit being 1 or 0 given a plaintext and a guessed part of the key. 
With an incorrect guess for part of the key the sum of all difference of averages along the trace 
would converge to zero while for a correct guess this converges to a non-zero value. 
The model that \citet{Kocher1999} used assumes that each individual bit of an intermediate value 
contributed to the power consumption of the device such that (with enough traces) it could be 
revealed. By extending the same idea to the power consumption of a register, we arrive at the 
Hamming weight model. 
This model states that the consumed power is 
proportional to the number of bits that are set~\cite{Mess00}. 

In another type of attack, called Correlation Power Analysis (CPA), 
\citet{BrierCO04} used the correlation coefficient as a side-channel 
distinguisher, i.e.\ the statistical method used for the key recovery. 
CPA 
allows an attacker to recover parts of a key that is used in a cipher by 
using a known plaintext attack: samples measured using a single probe are 
correlated against a synthetic power value that is generated from 
an intermediate value calculated for all values that a subkey can take. 
Commonly, the power model used for CPA is either Hamming weight or Hamming distance. 
In the Hamming distance model the consumed power is proportional to the number of different bits between 
two intermediate values. Such leakage can occur in practice when an intermediate value stored in a register is overwritten with another 
value. 

\subsection{Side-Channel Leakage Assessment} \label{s:scla}

Side-channel leakage assessment measures how vulnerable a device is to side-channel attacks. 
This cannot be 
an exhaustive assessment, as it is impossible to try all possible attacks on a device. 
However, such assessment is still 
valuable to the manufacturers of secure devices 
as they can benchmark a level of security of devices during the design and manufacturing process. 

In side-channel leakage assessment, the main question we try to answer is
whether the evaluated device shows significant leakage. Therefore, a device 
must show statistically
significant leakage to be classified as insecure. Standards such as 
International Standard ISO/IEC 17825:2016(E)~\cite{ISO17825} build on a 
methodology called
Test Vector Leakage Assessment (TVLA) that was initially presented by 
\citet{GoodwillJJR2011}.
The TVLA methodology uses Welch's \ttest~\cite{welch1947generalization} to 
detect statistical differences between sample distributions that are measured 
when the device processes different inputs. 
Two main test configurations are specified: the fixed vs.\ random
configuration and the fixed vs.\ fixed configuration. 
The reason for calculating such differences is that a protected cipher 
implementation 
should not be emitting any information that would let an evaluator 
differentiate the data it 
processes. If the calculated difference is statistically insignificant, the 
device is regarded as side-channel free in the context that it was tested on. 
It has been demonstrated that the results of \ttest{}s should not be 
misinterpreted as a single test that decides if a device is secure or 
not~\cite{Standaert18}. Specifically, the result only suggests that the \ttest failed to find leakage for the specific fixed inputs used and number of traces 
collected from the device.
For different fixed input values or for a greater number of traces significant 
leakage could be observable. 

Welch's \ttest defines a statistic called the $t$-value which is
calculated from the means ($\overline{X}_1$ and $\overline{X}_2$) and variances
($s^2_1$ and $s^2_2$) of distributions of collected traces at a given sample 
point. The $t$-statistic follows a Student's $t$-distribution with $v$ degrees 
of freedom. Given the number of samples in each distribution as $n_1$ and 
$n_2$, the $t$-value ($t$) and degree of freedom ($v$) are calculated as:
$$t = 
\frac{\overline{X}_1-\overline{X}_2}{\sqrt{\frac{s^2_1}{n_1}+\frac{s^2_2}{n_2}}}
\;\;\;\;\text{and}\;\;\;\;
v = \frac{ \left( \frac{s^2_1}{n_1} + \frac{s^2_2}{n_2} \right)^2 } { 
	\frac{  \left(\frac{s^2_1}{n_1}\right)^2 }{n_1-1}  +\frac{  
	\left(\frac{s^2_2}{n_2}\right)^2 
	}{n_2-1}  }.$$

The $t$-value 
tells us how significantly different the two 
distributions are. 
The hypothesis that these two distributions
originate from the same population needs to be rejected with some given 
level of confidence to show that they are different. 
This process is known as hypothesis testing. Hypothesis testing is the 
scientific method of 
ruling out hypotheses by rejecting them based on significant evidence against 
them.
The null hypothesis is the hypothesis that we assume to be correct by default. 
In TVLA, we assume 
that the device is not leaky until evidence, such as a significant $t$-value 
from the 
\ttest rejects it in favour of the alternative hypothesis. The alternative 
hypothesis here 
states that the two sample distributions are statistically different, which implies that the considered device is leaky. The 
threshold value of $4.5$ 
for significant leakage is chosen at a significance level ($\alpha$) of 
$0.00001$ under the assumptions that $s_1 \approx s_2$ and $n_1 \approx n_2$, 
such that the total number of traces ($n_1 + n_2$) is 
greater than 1,000~\cite{SchneiderM15}.

Using naive methods to compute $t$-values may result in numerical errors due to 
cancellation 
effects~\cite{ChanGL1983}. Schneider and Moradi~\cite{SchneiderM15} 
demonstrated computational improvements to overcome 
such issues. They also suggested online calculation of $t$-values in a single 
pass, speeding up
the calculations compared to more naive methods. Another common concern 
with the evaluation 
of masked implementations is the typical \ttest threshold of $4.5$. This value 
assumes a single 
independent \ttest. 
This threshold value is inadequate for large numbers of sample points, as the 
possibility of false positives (i.e.\ classification of leakage at sample 
points as significant when there is no actual leakage) increases due to the 
increased number of tests. \citet{DingZDSF17} discussed this further and 
proposes a method to increase the \ttest threshold according to the degree of 
freedom of the \ttest{}~\cite{SchneiderM15} and number of samples.

\subsection{Masking Techniques and Higher-Order Side-Channel Attacks}

To protect ciphers against side-channel attacks, a technique called 
masking~\cite{ChariJRR99,Mess00} has been proposed.
With masking, a sensitive intermediate value is split into multiple parts by using additional randomness. 
The additional random values are referred to as \emph{masks} and the values that the original 
value is split into are referred to as \emph{shares}. Depending on the order of masking, the number of 
shares increases. For example, in a $d$th order masking scheme there are $d+1$ shares in use. 
Only when all of the shares are combined, the original value can be revealed. 

Since masked implementations are secure against traditional attacks such as DPA 
and CPA, these attacks have been generalised to overcome masking by exploiting several 
leakage points simultaneously.
Generally, a $(d+1)$th order attack aims at breaking a $d$th order masked implementation. 
Such attacks first combine leakage occurring in $d+1$ intermediate 
operations and then a classical attack such as  CPA can be applied to recover the 
key. By increasing the number of shares, an implementer can increase the work 
that is required for an attack exponentially~\cite{ChariJRR99}.

In particular, \citet{IshaiSW03} show that a masked implementation with $d+1$ shares 
is secure against side-channel attacks in the $d$-probing 
model. 
The $d$-probing model considers an adversary that can only learn up to $d$ 
intermediate values that are produced during the cryptographic computation.
The model is usually considered a good approximation for modelling higher-order 
attacks.

Even though masking techniques can be theoretically secure against wide-range 
of side-channel attacks, many 
practical effects, such as glitches~\cite{MangardPG05,ChenHS09} or transitional 
effects~\cite{BalaschGGRS14} that can void the countermeasure and 
still leak the secret information. In such cases, $d$th order implementations 
reveal their secret values at $d$th or lower order analysis.
\citet{RenauldSVKF11} attribute this effect to breaching the
Independent Leakage Assumption (ILA), which states that all related shares 
should be manipulated independently. 
Even though the ILA is assumed in theoretical cryptography, in reality this assumption does not hold due to 
the way that modern computers work. For example, to increase performance and reduce manufacturing 
costs, modern CPUs reuse many of their internal components without resetting or wiping them. 
\citet{BalaschGGRS14} demonstrated that transitional effects can be destructive to masked implementations as they halve 
the effective order of the analysis required when the leakage is modelled with a Hamming-distance
leakage model. 
In the Hamming weight 
model, only a single intermediate value is considered as a sample 
at a sample point. 
In contrast, the Hamming distance 
model uses the bit difference between 
two intermediate values for a sample. 

To measure the effectiveness of an implemented countermeasure such as masking, one needs to look 
into the leakage assessment of cryptographic devices.

\subsection{Higher-Order Side-Channel Leakage Assessment}\label{s:bivariate}
As discussed before, increasing the number of shares 
significantly increases the attack complexity, and information from multiple samples needs to be combined to reveal leakage of higher-order implementations. 

In contrast to univariate analysis, where each sample point is 
analysed independently of other points, higher-order analysis takes into 
account the joint leakage of two or more sample points. 
This is similar to using multiple probes with respect to the model of \citet{IshaiSW03}. 
A combination function is typically used to combine mean-centered samples, and
leakage assessment is then carried on the resulting combination. 
Following \citet{ProuffRB09},
we choose the `product of samples' combination function,
In case of multivariate 
\ttest{}s, the result of the combination is used as input to a first-order 
\ttest{} and analysed similar to the univariate case~\cite{SchneiderM15}. 

Let us consider a set of $n$ side-channel measurements $T_i$, 
$0\leq i <n$, which are known as \emph{traces}. 
Each trace contains $m$ sample points denoted as $t^{(j)}_i$, for $0\leq j <m$ with sample means denoted by $\mu^{(j)}$, 
Then the mean centered product of a given subset of sample points $\mathcal{J}$, is given by: 
\begin{equation}\label{eq:biv}
C_i = \prod_{j \in \mathcal{J} } \left(t^{(j)}_i - \mu^{(j)}\right).
\end{equation}
When $|\mathcal{J}| = 2$ the combinations generated are called  \emph{bivariate}
and when $|\mathcal{J}| = 3$ they are called
\emph{trivariate}. 

Usually we need to consider all possible subsets $\mathcal{J}$ in a given trace 
$T_i=t^{(0)}_i, \dots t^{(m-1)}_i$ to detect the leakage using \ttest{}. 
Therefore, the complexity increase from using this approach higher-order leakage assessment is by a factor of $\binom{m}{|\mathcal{J}|}$,
which is exponential for small values of $|\mathcal{J}|$.

\subsection{Leakage Emulators and Automatic Countermeasures}

Due to the high costs associated with evaluations that use real devices, implementers of 
cryptographic code are inclined to use emulators to determine leakage of a 
device~\cite{BuhanBYS21}. The first use of such an emulator is evidenced within the PINPAS project~\cite{HartogVVVW03}, which had as the goal to 
emulate power analysis leakage in Java cards. 

The most accurate method to emulate leakage is circuit-level emulation. 
While accurate, it is also very slow due to the very realistic reproduction of internal effects. Earlier 
generations that emulate leakage for software implementations used the cipher source code written 
in an high-level language~\cite{Veshchikov14, Reparaz16}. However, such implementations are 
inadequate for detecting leakage stemming from breaches of the ILA. In addition, compilation
can also introduce breaches of ILA.
Consequently, recent leakage emulators tend to use machine code as input
rather than high-level source code~\cite{McCannOW17,CorreGD18,Papagiannopoulos17}. 
\citet{Papagiannopoulos17} developed an automated methodology for detecting violations 
of the ILA in AVR assembly. They investigate the effects of the 
ILA violations on an AVR microcontroller, ATMega163. By enforcing the ILA, the authors produce a 
first-order secure S-box for the RECTANGLE~\cite{ZhangBLR0V15} cipher.

With Coco~\cite{GigerlHPMB20}, it is possible to formally verify a masked implementation 
down to the gate-level when the netlist of the CPU is available. A major difference between the 
construction of other leakage emulators and Coco
is that Coco uses a software tool 
called Verilator to convert Verilog hardware descriptions of the CPU into C++. This enables 
the construction of a detailed emulator and offers fine-grained information about the execution. It 
collects power information for each gate and then uses a SAT-solver to find the leaky gates. 
While Coco finds the exact gates that are contributing to the leakage, it does not provide an
automated fixing mechanism.

\citet{McCannOW17} demonstrated an emulator named \elmo that emulates leakage based on machine 
instructions. The emulation uses a statistical model that is profiled using real 
traces. This makes it specific for the device it was profiled on.
\elmo currently supports ARM Cortex-M0 and ARM Cortex-M4 processors.

The recently introduced \rosita~\cite{SSB19} aims to automate the process of producing 
masked first-order implementations. \rosita uses leakage information from an improved version of 
\elmo~\cite{McCannOW17}, which the authors call \elmos, to emulate 
the power consumption of the target device running the software.
It then uses TVLA~\cite{GoodwillJJR2011} on the emulated traces to detect 
instructions that leak information.
When leakage is detected, \rosita performs root cause analysis to identify 
the cause of the leakage.
Specifically, it performs \ttest analysis on emulated traces of each of the 
components of the \elmos model, identifying a components that show evidence of leakage.
Based on the root 
cause, \rosita applies rewrite rules, modifying the cipher code with the aim 
of eliminating the leakage. The process repeats until either no more rules can 
apply or no leakage is evident.

Similarly, \citet{GaoMPP20} have demonstrated an Instruction Set Extension (ISE) to 
RISC-V Instruction Set Architecture (ISA). The ISE guarantees that internal states that cause 
leakage are cleared acting as a barrier instruction when used in sensitive 
programs.

\subsection{Testing for Statistical Equivalence of Distributions}
\label{sec:tsop}
In \cref{s:rootcause} we 
use a statistical equivalence test during root-cause 
analysis to determine which parts of the code contribute to the leakage;
In this section, we 
describe the statistical method we use for equivalence testing.

The aim of statistical equivalence tests is to determine how probable it is that two sampled distributions 
originate from the same population. Observe that this is the opposite of what 
Welch's \ttest offers. The null hypothesis of an equivalence test is that the two 
distributions are different and we expect to reject it in favour 
of the alternative hypothesis which states that the distributions are the same 
with a given significance level. One such equivalence test is the Two One Sided 
\ttest (TOST)~\cite{Sch87,PardoScott2013EaNT}.
As the name indicates, TOST uses two one-sided \ttest{}s to test whether the two distributions are equivalent.
TOST is a 
parameterised 
test that requires a lower bound and upper bound for the mean difference of the 
two distributions under test as parameters. Two individual $t$-tests determine 
whether the mean difference is lower than the upper bound and whether it is 
higher than the lower bound with a given level of significance ($\alpha$). 
Passing both \ttest{}s indicates that the mean difference is between the lower 
and upper bounds with the given significance level.
 
However, TOST in its original form has a limitation when it comes to the 
evaluation of the mean differences of two distributions: when these mean 
differences 
are close or equal to the boundary values, the TOST concludes that the distributions are not 
equivalent. This happens due to the $t$-value of the individual \ttest{}s resulting in 
values closer to $0$ when the mean differences are close to boundary values. In the paradigm where 
TOST is commonly used (e.g. in drug test trials), the boundaries are regarded as the worst values 
that the mean difference can get. However, in equivalence testing for 
engineering, we expect a test which  accepts boundary values and also the 
values which are closer to the boundaries. 

To mitigate this limitation, \citet{PardoScott2013EaNT} proposed the 
following formulas that compute new boundaries ($\overline{X}_H$ and $\overline{X}_L$) given a 
target mean difference ($\mu$), where $s$ and $n$ are standard deviation and cardinality of the 
mean differences distribution. $t_\alpha$ is the one sided \ttest value at a 
significance value of 
$\alpha$. 

Selecting a critical region with $\alpha$ significance level such that 
$\overline{X}_H$ is high\textbf{}er 
than $\mu$ is given as
\begin{equation} \label{eq:boundup}
\begin{split}
\overline{X}_H & = \mu + t_{\alpha}\frac{s}{\sqrt{n}}
\end{split}
\end{equation}
and 
such that $\overline{X}_L$ is lower than $\mu$ is given as 
\begin{equation} \label{eq:boundlow}
\begin{split}
\overline{X}_L & = \mu - t_{\alpha}\frac{s}{\sqrt{n}}.
\end{split}
\end{equation}

Using the confidence interval of $\overline{X}_{L}$ and 
$\overline{X}_{H}$ instead of having arbitrarily defined values for mean 
difference boundaries overcomes the above-stated limitation.

\section{\horosita Design} \label{sec:horosita}

Past solutions that aim to automate leakage 
detection~\cite{Veshchikov14,McCannOW17,HartogVVVW03,Papagiannopoulos17,SehatbakhshYZP20}
and correction~\cite{SSB19} focus on first-order leakage. As the 
security of cipher implementations can be increased by employing more shares in 
their masking schemes, there is a  need for emulators and countermeasures that can work 
with multivariate leakage. In this section 
we describe how \horosita, our solution for this need,  works.
We outline the main challenges in performing higher-order analysis and proceed 
to describe our approaches for addressing these challenges.

\subsection{Challenges for Higher-Order Analysis}\label{s:challenges}
The core extension required for \horosita to support higher-order leakage 
detection and mitigation is support for multivariate analysis.
Specifically, instead of looking for instructions that show indication of leakage, we need to look for combinations of instructions that \emph{together} show indication of leakage.

\looseness=-1
\citet{SchneiderM15} suggest a methodology for performing multivariate analysis.
Their approach is to generate artificial, multivariate traces from the original 
univariate traces.
For that, the original traces are first preprocessed by calculating the average value for each sample point and subtracting the average from the corresponding point in each trace.
As \cref{eq:biv} shows, 
Each point in an artificial trace represents a tuple 
of points in the original trace, where the value associated with the artificial 
point is the product of the values for the corresponding points in the 
original trace.

Our approach for performing higher-order analysis is to replace the use of TVLA 
in \rosita with the Schneider-Moradi methodology.
However, while seemingly straightforward, the approach raises significant challenges.

\challenge{c:stats}{Statistical confidence with multivariate traces}

The artificial $d$-variate traces have an artificial sample for each combination 
of $d$ samples in the original traces. 
Consequently, the length of the multivariate traces grows exponentially with 
the length of the original traces.
For traces of length $n$, the multivariate traces have a length of ${n 
\choose d}$ samples.

The de-facto standard statistical test used in TVLA is to reject the null 
hypothesis, i.e.\ report leakage, when the absolute value of the \ttest is 
above a threshold of 4.5, achieving a significance level of 0.00001.
This test, however, fails to account for the multiple comparisons performed in  TVLA, where a statistical test is performed independently on each sample point.
For a small number of points, the effect of multiple comparisons is negligible.
When the trace length increases, multiple comparisons result in false positives, showing leakage where no leakage exists.

\challenge{c:datasize}{Increased data size}
Another issue with multivariate analysis is the increase in the volume of data 
that needs to be processed compared with first-order univariate analysis.
Three factors contribute to this increase.
First, due to the effects of noise, the number of traces required for statistical analysis grows exponentially with the order of analysis~\cite{ChariJRR99}. 
Secondly, as discussed, the artificial multivariate traces are significantly 
longer than the original traces.
Thirdly, to increase the statistical confidence while handling \cref{c:stats} without missing leakage we need to increase the number of traces we process.

Because \horosita repeatedly evaluates implementations, there is a need for efficient methods for handling the increased amount of data with minimal impact on analysis time.

\challenge{c:rootcause}{Multivariate root-cause analysis}
The third challenge we face relates to performing the root-cause analysis. 
\rosita performs the analysis using a \ttest on each of the \elmos model components.
Such an approach can detect univariate leakage.  
However, detecting multivariate leakage necessitates evaluating combinations of 
components.
A brute-force approach that evaluates a \ttest statistics on every combination  of components is computationally expensive, particularly considering the increased number of traces, as described in \cref{c:datasize}. 
Thus, new techniques for root-cause analysis are required.

We now discuss how \horosita addresses these challenges.

\subsection{Achieving Statistical Confidence}\label{s:statconf}
\begin{figure}[htb]
	\begin{center}
		\includegraphics[scale=0.75]{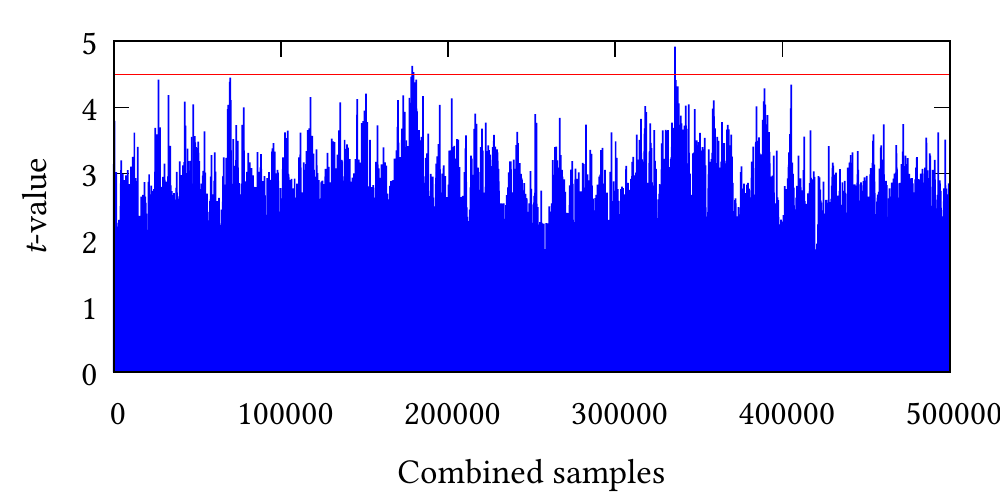}
	\end{center}
	\caption{A \ttest threshold value of 4.5 for a bivariate analysis with 1000 samples with all 
	inputs being random. }
	\label{f:thres}
\end{figure}

As discussed, \cref{c:stats} is that, due to the exponential increase in the number of sample point  per 
trace, the \ttest threshold of 4.5 is no longer appropriate. This mostly affects the traces 
collected from the physical experiment where we collect longer traces (10 times more 
samples) to reduce the effects of noise.
To demonstrate the false positives we collect 500,000 bivariate traces of a three-share implementation of 
Xoodoo (further described in \cref{s:xoodoo}) running on a STM32F030 Discovery 
evaluation board, where all inputs are drawn uniformly at random. The 
experiment setup we used is described in~\cref{s:expsetup}.
We then split these arbitrarily into two populations, and perform a bivariate \ttest analysis,  comparing these populations with a threshold of 4.5.
As \cref{f:thres} shows, despite the populations being sampled from the same 
distribution, several false positives are present.

For engineers, these false positives are typically of low impact.
Experienced engineers can typically identify false positives, e.g.\ by observing the context.
Alternatively, repeating the test can confirm true positives.

Automatic tools, such as \horosita, do not have the experience or the insight, and must rely on   
statistical tools for handling false positives. If \horosita is used with long code segments these 
false positives will also be present in its leakage analysis. Therefore, in \horosita, we adopt the 
approach of \citet{DingZDSF17}, who propose increasing the threshold to reduce the
probability of false positives.
Specifically, Ding et al.\ provide a formula to calculate the threshold given 
the number of samples and a desired significance level $\alpha$. 
We apply the formula to the length of the bivariate trace aiming for a 
significance level of 0.00001.
This ensures that the probability of a false positive error is less than .001\%, which we consider negligible.
For the traces in \cref{f:thres} we would use a threshold of 6.71, which is clearly above the largest peak in the figure.
Hence, at this threshold, the analysis does not indicate any leakage, which is expected considering that the two populations are drawn from the same distribution.

\subsection{Handling Large Datasets}\label{s:size}
As mentioned, several aspects of multivariate analysis result in a significant increase in the size of data that \horosita needs to process.
First, for given mean and variance, the Welch $t$-value grows linearly with the square root of the size of the population.
Consequently, when increasing the threshold we need a quadratic increase in the number of traces to achieve the same detection sensitivity.
Second, the length of the multivariate artificial traces is several orders of magnitude longer than the original univariate traces.
Third, due to the effects of noise, detecting higher-order leakage is inherently harder then detecting first-order leakage.
The combined effect of these changes is that the amount of data that \horosita needs to process is 
several orders of magnitude larger than that of \rosita. 
When evaluating the final version 
of code produced by \horosita on real hardware, the same issue gets even more apparent because we 
use longer traces as mentioned in \cref{s:statconf}.

While we are aware of works that have performed analyses at scales similar and even larger than our 
work~\cite{CnuddeBRNN15,CnuddeRBNNR16}, we could not find public tools that perform such analyses, 
or even performance figures for the analysis.
Free tools such as Jlsca\footnote{\url{https://github.com/Riscure/Jlsca}}, Scared%
\footnote{\url{https://gitlab.com/eshard/scared}}, SCALib%
\footnote{\url{https://github.com/simple-crypto/SCALib}}  seem to only offer limited capabilities.
To address this challenge we developed analysis tools from the ground up.
Our analysis tools avoid the overhead of storing the artificial traces (i.e.\,multivariate 
combinations) by calculating them on the fly.
The tools are multithreaded, allowing a significant speed-up,
and the data is divided point-wise between the threads, so that each thread only accesses a limited subset of the original traces' samples.

We acknowledge that the approach is fairly straightforward, but we believe that the contribution is 
important for practical future research into bivariate analysis.

\subsection{Multivariate Root-Cause Analysis}\label{s:rootcause}
The third challenge for \horosita is performing root-cause analysis on 
multivariate traces.
The \elmos linear regression model consists of 28 term components, each 
modelling a different micro-architectural effect.
When \rosita performs univariate root-cause analysis, it calculates the Welch 
$t$-value for each component separately, where the leaky components are 
identified by observing significant $t$-values.

While this approach works well for univariate leakage detection, adapting it to 
multivariate leakage is not trivial. The main reason is that, in multivariate 
analysis, there is no single cause for leakage.

As shown by \cref{eq:biv}, a multivariate sample point is a combination of many 
samples in the original trace. In \elmos, each of the original samples is 
calculated from the sum of 28 model components. 
Searching for a combination of $d$ samples using a method similar to the one 
used for univariate evaluation would require evaluating $28^d$ combinations.
Even for the bivariate case of $d=2$, the process is very inefficient with 
the 
large number of traces that need to be processed due to increase of 
order~\cite{ChariJRR99}.

To avoid searching the whole space of pairs of model components, \horosita uses two new methods for finding the components that contribute to the leak.
The \emph{component elimination} method tests whether removing a model component removes the leakage. 
While efficient, this approach may sometimes fail.
In the case of such a failure, \horosita reverts to a \emph{Monte-Carlo method}, which tests random combinations of components looking for evidence of component leakage.
We refrain from using the Monte-Carlo method by default due to its inefficiency and the instability inherent in a randomised process.  
We now describe these two methods in detail.

\parhead{Component Elimination}
The basic idea behind the component elimination method is to identify
components that contribute to the leakage by removing one component at 
a time from the multivariate sample combination function (which is shown 
in~\cref{eq:biv}); we then evaluate the combination with removed component for absence of leakage. 
If the removal of a 
component leads to the absence of leakage at a previously leaky point, this means 
that the removed component contributed to the leakage. When this process 
ends, \horosita has a set of components 
that contribute to the leakage. \horosita can now apply fixes
using the approach of \rosita.

More specifically, component elimination consists of the following steps.
First, each component value of \elmos is recorded with the component index, 
sample index, and the trace index. There exists 28 different components in 
\elmos, the sample index is the array index of the instruction when the 
emulated code segment is unrolled into individual instructions. The trace index 
is a number identifying each run of the fixed vs.\ random test. All of these 
values are stored in a 3D matrix that is denoted by $\mat{L}$. 

Second, the multivariate leaky points for the implementation are found by 
running 
the 
\ttest{}
on the final power value of~\elmos{} generated while running 
the code segment in a fixed vs.\ random input configuration.

Finally, \cref{alg:flc} is run to find the leaky components at the leaky points 
recognised in the previous step. The two utility 
functions that are used by \cref{alg:flc} are Normalised Product of 
Samples (\textsc{NPS}) and \textsc{NotLeaky}. The 
first function, \textsc{NPS} returns the combined traces for a given set of 
sample points and a given set of components. In a nutshell, \textsc{NPS} 
returns the results of \cref{eq:biv} for an arbitrary set of components and 
an arbitrary set of sample points. As the name suggests, the \textsc{NotLeaky} 
function differentiates between trace sets which are significantly similar and 
ones which are not. 
\textsc{NotLeaky} requires an additional run of the code segment 
with 
all random input configuration instead of a fixed vs.\ random input 
configuration. This run collects information required to calculate the mean 
differences and variances required by TOST. 

\begin{algorithm}
  {\small
  \begin{description}[style=unboxed,leftmargin=0.8cm]
    \item $\mat{L}$: A 3D matrix with component values for \elmos organised by 
    trace index, sample index and component index.
    \item $\mathcal{S}$: Set of $d$ sample points that participate in the 
    leakage.
    \item $\mathcal{C}$: Set of all components that are in \elmos{}.
    \item \textsc{NPS}$\left(\mat{L}, \mathcal{S}, \mathcal{C}\right)$:
    Normalised Product of Samples. Returns 
    the normalised product of the power
    samples from reduced models which only contain a given set of 
    components 
    ($\mathcal{C}$) at some given sample points ($\mathcal{S}$) from a 3D matrix that holds component 
    samples 
    ($\mat{L}$).
    \item \textsc{NotLeaky$\left(\mat{Y}\right)$}: Determine the absence 
    of leakage using TOST.
    \item $\odot$: Elementwise multiplication operator.
  \end{description}

	\begin{algorithmic}[1]
		\Function{FLC}{$\mat{L}, \mathcal{S}, \mathcal{C}$}
		\State {$r \gets \{\}$}
		\For{$s \in \mathcal{S}$}
		\State { $ \mat{x} \gets \Call{NPS}{\mat{L}, \mathcal{S} \setminus s, 
				\mathcal{C}} $ }
		\For{$t \in \mathcal{C}$}
		\State $u \gets \mathcal{C} \setminus t$
		\State $\mat{y} \gets \Call{NPS}{\mat{L}, s, u}$
		\State $\mat{z} = \mat{y} \odot \mat{x} $
		\If{$\Call{NotLeaky}{\mat{z}}$}
		\State {$r \gets r \cup \{(s,t)\}$}
		\EndIf
		\EndFor
		\EndFor
		\State \Return {$r$}
		\EndFunction
	\end{algorithmic}
  }
	\caption{Find Leaky Components} 
	\label{alg:flc}
\end{algorithm}

While the component elimination method is efficient, it may sometimes fail.
For example, if multiple model components leak the same share, removing any one 
of these components will not eliminate the leak.
Similarly, TOST may fail to demonstrate the equivalence of the two 
distribution even when 
removing a model component eliminates the leak. 

\parhead{The Monte-Carlo Method}
In the Monte-Carlo approach we run a preset number of random experiments where,
in each experiment, we select a random subset of the model components, and perform
the \ttest on hypothetical power traces with only the selected components.
For each component, we keep track of the number of random experiments it participates in and how many of those experiments indicate significant leakage. After we repeat the experiment a preset number of times, we arrive at a subset of components that contribute significantly more to the leakage.

\begin{figure}[htb]
	\begin{center}
		\includegraphics[scale=0.75]{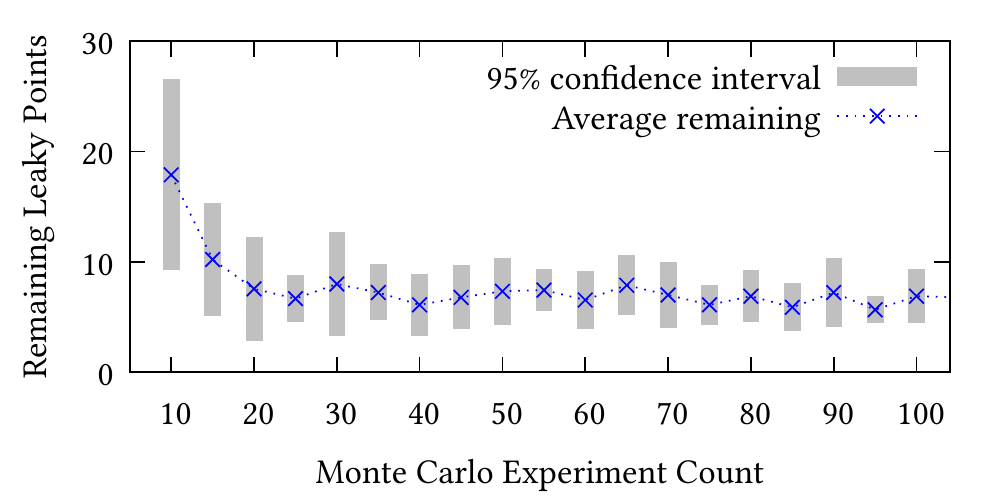}
	\end{center}
	\caption{Effectiveness in removing leakage of Monte Carlo method for 
	increasing number of 
		experiments}
	\label{f:monte}
\end{figure}

To select the preset number of random experiments we first performed an 
analysis for a code segment from Xoodoo cipher (shown in \cref{l:xoodoo3}) only by using Monte Carlo method to detect 
and remove leakage. 
We use the chosen preset number in all our subsequent experiments.
We first gathered 100,000 traces from this cipher implementation and performed the initial leakage analysis. Initially, it had 45 total leakage points. \cref{f:monte} shows the reduction of remaining leaky points as we gradually increase the number of Monte Carlo 
experiments starting from 10. \cref{f:monte} shows that increasing number of experiments improves detection of root causes, but after about 30 experiments the reduction of leakage is nearly constant\footnote{We have noticed no further leakage reduction even for 1000 experiments.}. 
Therefore, we decided to settle at using 50, slightly more than 30 for the sake of certainty, as the preset experiment count for 
our experiments.

Observe that both component elimitation and the Monte-Carlo method are independent of the security order. 
We evaluate both methods in the second and third order in \cref{s:evaluation}. 

In our experiments we find that we need to fallback to  the Monte-Carlo method
in four out of 16 root-cause detections in Xoodoo, in 70 out of 262 in 
\present, and in one out of 15 in the Boolean-to-arithmetic conversion algorithm.

\subsection{Code Rewrite}
After finding the root cause of the leakage, \horosita selects the code-rewrite 
rule that best match the detected root cause using the code-rewrite engine of
\rosita.

In a nutshell, \rosita reserves the register \texttt{r7}, which it initialises with a random value.
When an unintended interaction is detected, the code rewrite engine inserts instructions that use \texttt{r7} to eliminate the interaction.
For example, when the detected interaction is caused by a pipeline register that is updated by two consecutive instructions, \rosita inserts the instruction \texttt{mov r7, r7}, to buffer between the interacting instructions.
Similarly, when the leakage is from an interaction with the memory subsystem, \rosita inserts the pair of instructions \texttt{push \{r7\}} followed by \texttt{pop \{r7\}}, which wipes the internal state of the memory pipeline.
Many other fine-grained fixes are used to erase internal state set by other 
instructions (i.e. instructions related to ALU's operations). 

Observe that we use the same code rewriting engine that was initially designed 
for fixing univariate leakage. We find that it is usable as is to also fix 
multivariate leakage becasue the output of our root-cause detection algorithm matches the format of the original \rosita output.
The downside of reusing the code rewriting engine is that we may miss opportunities for addressing multiple leaks with a single fix.
We leave optimising the code-rewrite engine to future work.

\section{Evaluation}\label{s:evaluation}
In this section we evaluate the effectiveness of \horosita in eliminating 
leakage. First, we describe our physical experiment setup. Second, we present 
toy Boolean masked examples of second and third order where \horosita fixes a 
single leaky point. Third, we present the evaluation results of \horosita's 
emulation process and root cause detection. Finally, we demonstrate 
effectiveness of \horosita on practical code segments 
implemented with 3-shares. 
Due to practical reasons, we limit the discussion to second and third order.
We note that \horosita can detect and apply fixes at any order.

\subsection{Experimental Setup} \label{s:expsetup}

\begin{figure}[htb]
	\begin{center}
  \includegraphics[scale=0.3]{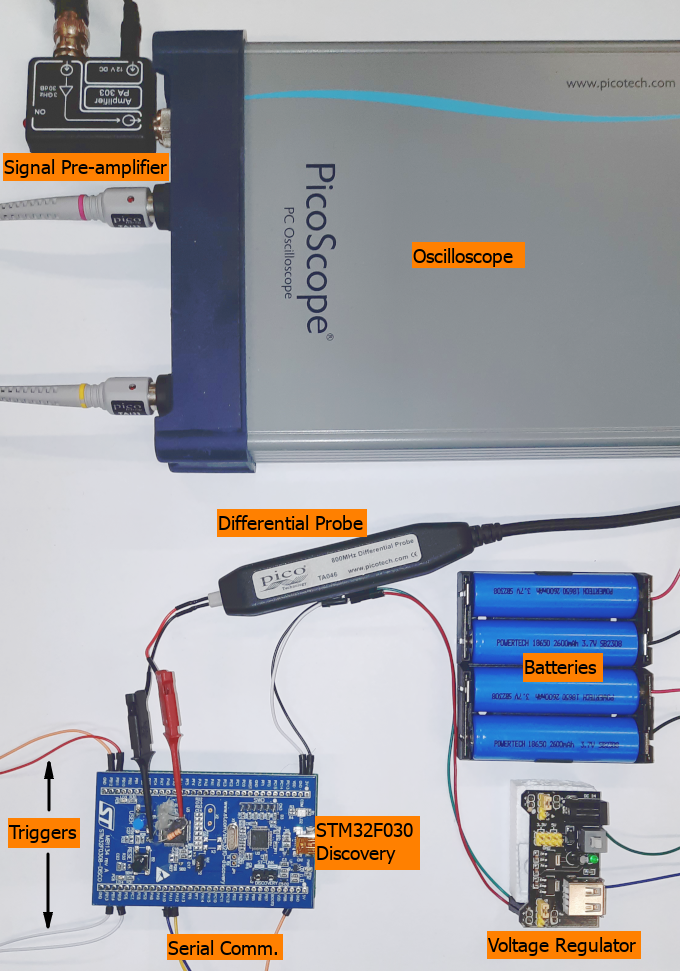}
	\end{center}\vspace{-2mm}
	\caption{Measurement setup}
	\label{f:expsetup}
\end{figure}

Our experimental hardware setup is depicted in \cref{f:expsetup}.
For evaluation we use the STM32F030 Discovery evaluation board by ST Microelectronics, which features an ARM Cortex-M0 based on STM32F030R8T6 System-on-Chip (SoC), running at 8\,MHz.
To avoid switching noise, we power the evaluation board with batteries instead of a mains-connected power supply.

To measure the power consumption of the evaluation board, we introduce a shunt resistor across one of its power terminals. 
We measure the voltage drop across the shunt resistor with a PicoScope 6404D oscilloscope, configured at a sampling rate of 78.125\,MHz (12.8\,ns sample interval) which translates to roughly 9.77 samples per clock cycle. 
The voltage is measured with a PicoTechnology TA 046 differential probe connected to the 
oscilloscope via a Langer PA 303 preamplifier.

We use two of the  I/O pins of the board to trigger the 
acquisition.
One indicates trace start and the other indicates the end.
To increase signal stability, interrupts are disabled for the duration of each trace,
using \texttt{\_\_disable\_irq()} before the start trigger and \texttt{\_\_enable\_irq()} after the end trigger.

To orchestrate the experiment, we used a PC with a serial connection to our device under test. 
The PC controls all aspects of the experiment, and in particular it selects the type of the experiment (i.e.\ fixed vs.\ random) and the randomness used.
The tested device is oblivious to the type of experiment and uses the inputs 
received from the PC. 
To reduce the communication overhead the PC uses bulk transfer to send the inputs for multiple successive experiments, which the device executes sequentially.

We post-processed the traces 
to improve signal quality. 
Firstly, we aligned the traces statically using a correlation-based alignment, reducing sample drift.
We then used a highpass filter to remove frequencies below 400\,KHz.
Before filtering, the signal was zero padded to 
avoid introduction of transients~\cite{Smith97DSP}.

\subsection{Evaluation of second and third-order Boolean masked toy example}
\vspace{-.5cm}
\begin{lstlisting}[
  numbers=left,
  numberstyle=\scriptsize,
  basicstyle=\footnotesize\ttfamily,
  keepspaces,
  xleftmargin=2em,
  caption=A Toy Example (second order),
  label=l:example,
  escapechar=|]
  ; nop padding
ldrb r4, [r1] |\label{line:synth1}|
push {r7}     |\label{line:push}|
pop {r7}   |\label{line:pop}|
  ; nop padding |\label{line:pad}|
ldrb r5, [r2] |\label{line:synth2}|
ldrb r6, [r3] |\label{line:synth3}|
  ; nop padding
\end{lstlisting}
Before we evaluate \horosita on real-world software examples, we demonstrate its effectiveness on a toy example, shown in \cref{l:example}.
The code presents a typical operation in second-order protected implementation 
that uses Boolean masking.
Specifically, it assumes that registers \texttt{r1}, \texttt{r2}, and \texttt{r3} contain the addresses of three shares that represent a secret value.
The code uses three \texttt{ldrb} instructions to load the masked value into three registers, \texttt{r4}, \texttt{r5}, and \texttt{r6}.
We note that the code is nominally second-order secure, because all instructions process at most one share of the secret.
However, as we see below, unintended interactions between the load instructions at \cref{line:synth2,line:synth3} result in second-order leakage.

To avoid first-order leakage through a combination of the three load instructions, we separated the first load (\cref{line:synth1}) from the rest of the code.
We added the \texttt{push} and \texttt{pop} instructions in \cref{line:push,line:pop} to remove interactions between the first load and the following two loads. (See \citet{SSB19} for details.)
We further added sequences of nine \texttt{nop} instructions (concretely, 
\texttt{mov r7, r7}) to avoid unintended interaction through the processor's 
pipeline and to achieve a clear temporal separation between the loads.
Last, we add short sequences of \texttt{nop} instructions around the code to create a temporal separation between the measured code and the triggers.

\begin{figure}[htb]
	\begin{subfigure}[t]{\linewidth}
  \begin{center}
	\includegraphics[scale=0.75]{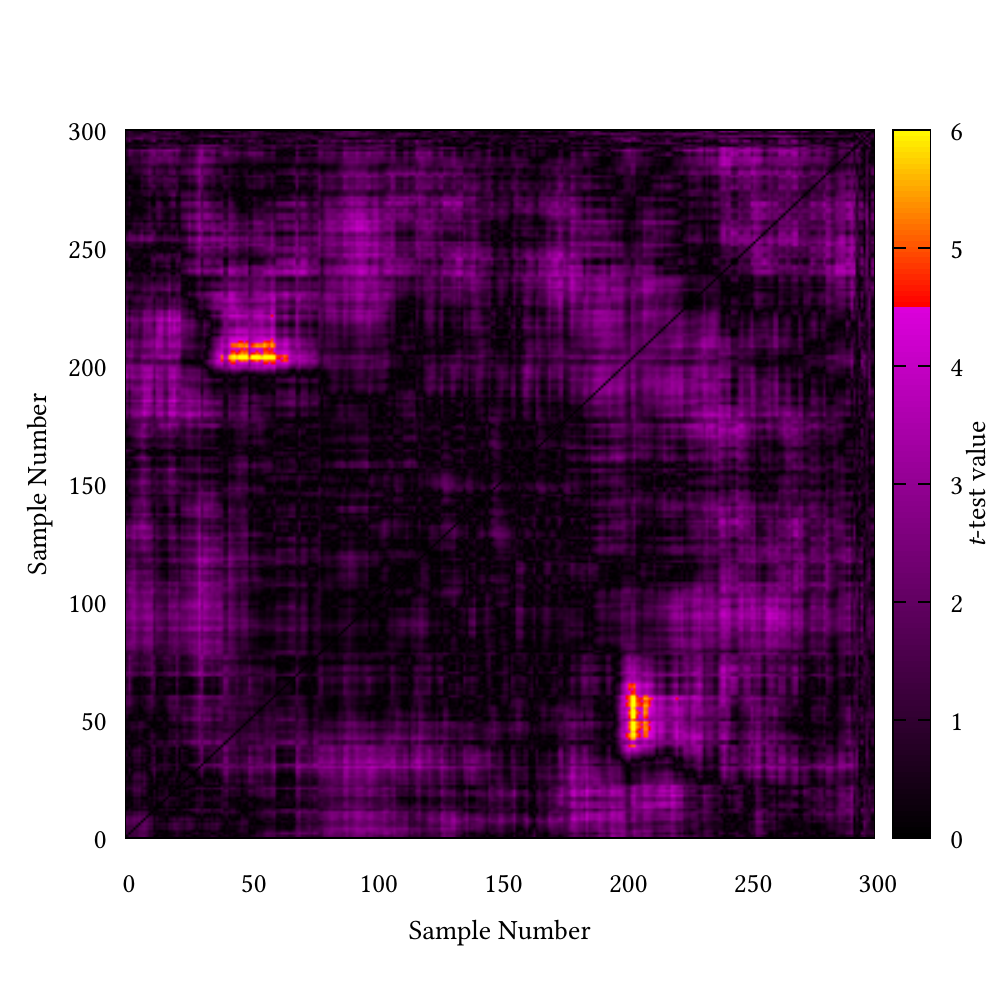}
	  \caption{Before applying code fixes. Leakage is visible around 
	  coordinates (50,200) and (200,50). $t$-value peak: 6.86.\label{f:synthetic_before}}
   \end{center}
	\end{subfigure}
	\begin{subfigure}[t]{\linewidth}
  \begin{center}
	\includegraphics[scale=0.75]{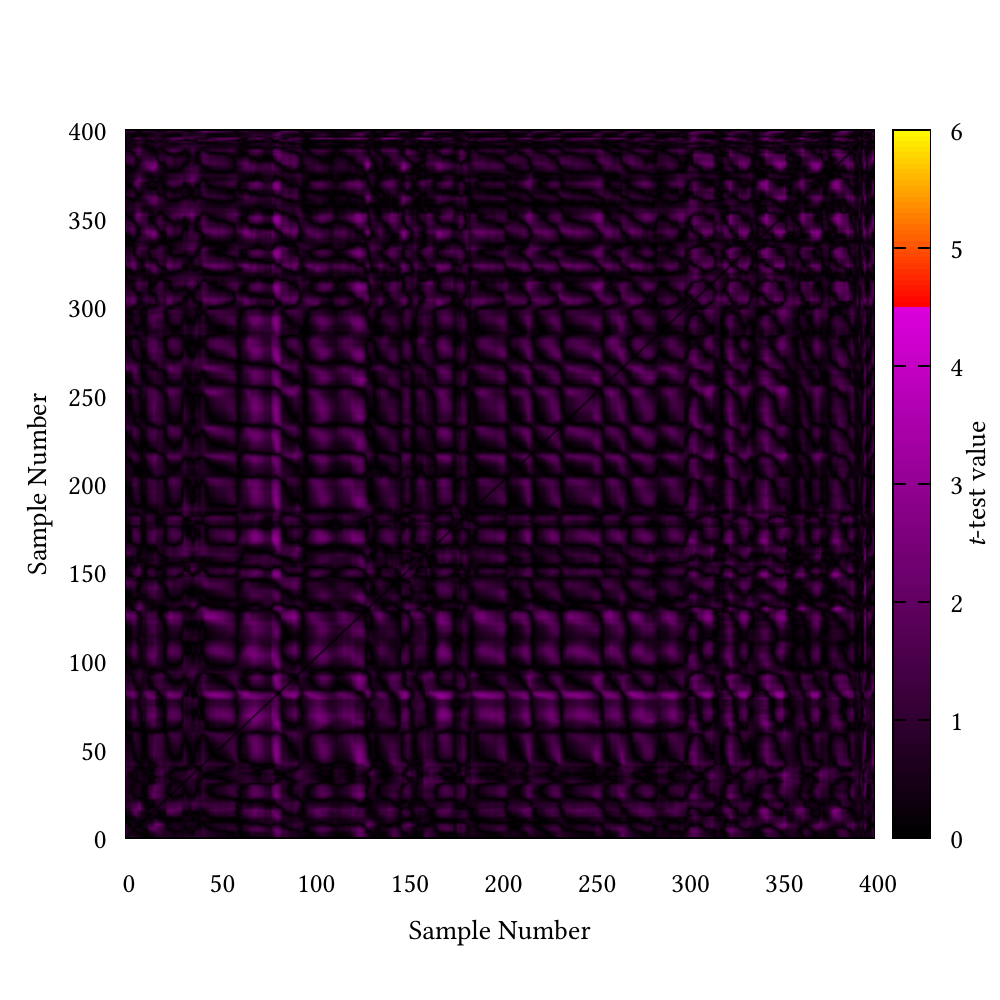}
	\caption{After applying code fixes no leakage is present ($t$-value peak: 3.15)\label{f:synthetic_after}}
  \end{center}
	\end{subfigure}
	\caption{Evaluating a toy example.}
	\label{f:synthetic}
\end{figure}

\begin{figure}[htb]
	\begin{subfigure}[t]{\linewidth}
  \begin{center}
		\includegraphics[scale=0.75]{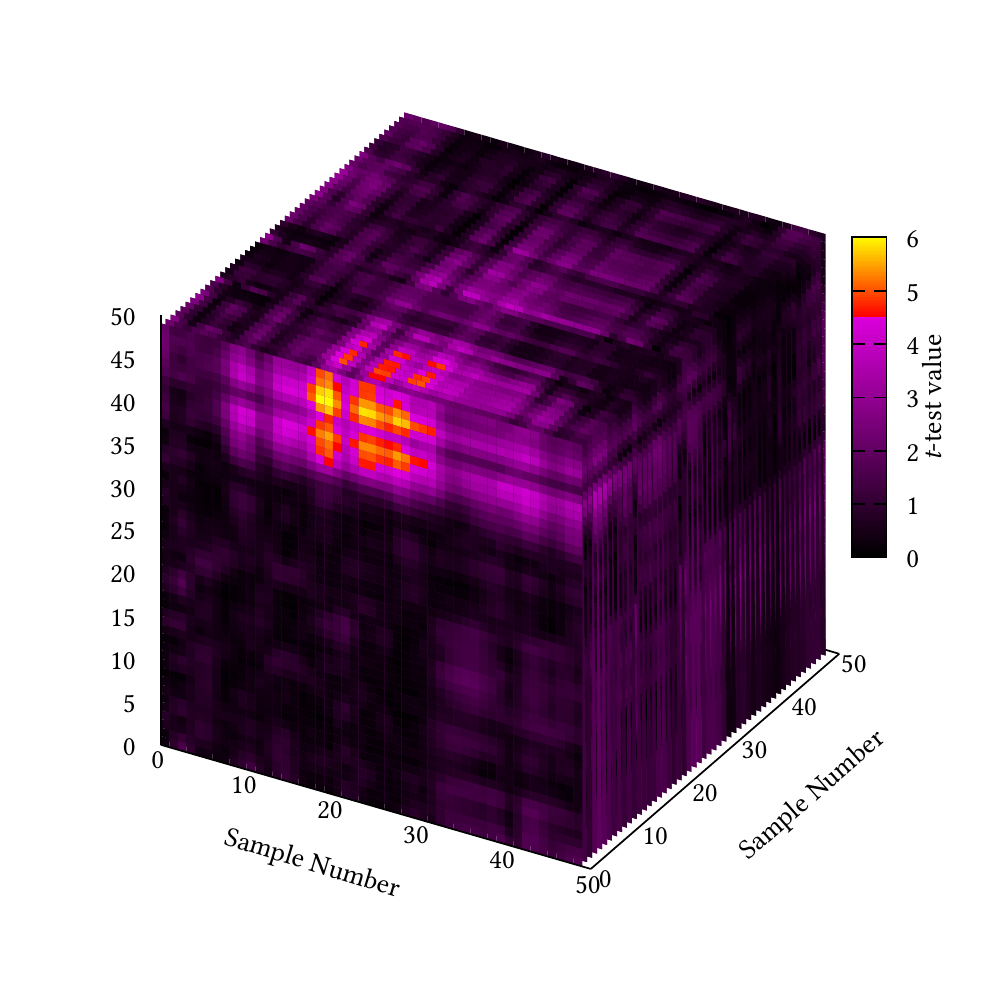}
  \end{center}
		\caption{Before applying code fixes. $t$-value peak of 6.85 at (42,28,7)\label{f:synthetic4_before}}
	\end{subfigure}
	\begin{subfigure}[t]{\linewidth}
  \begin{center}
		\includegraphics[scale=0.75]{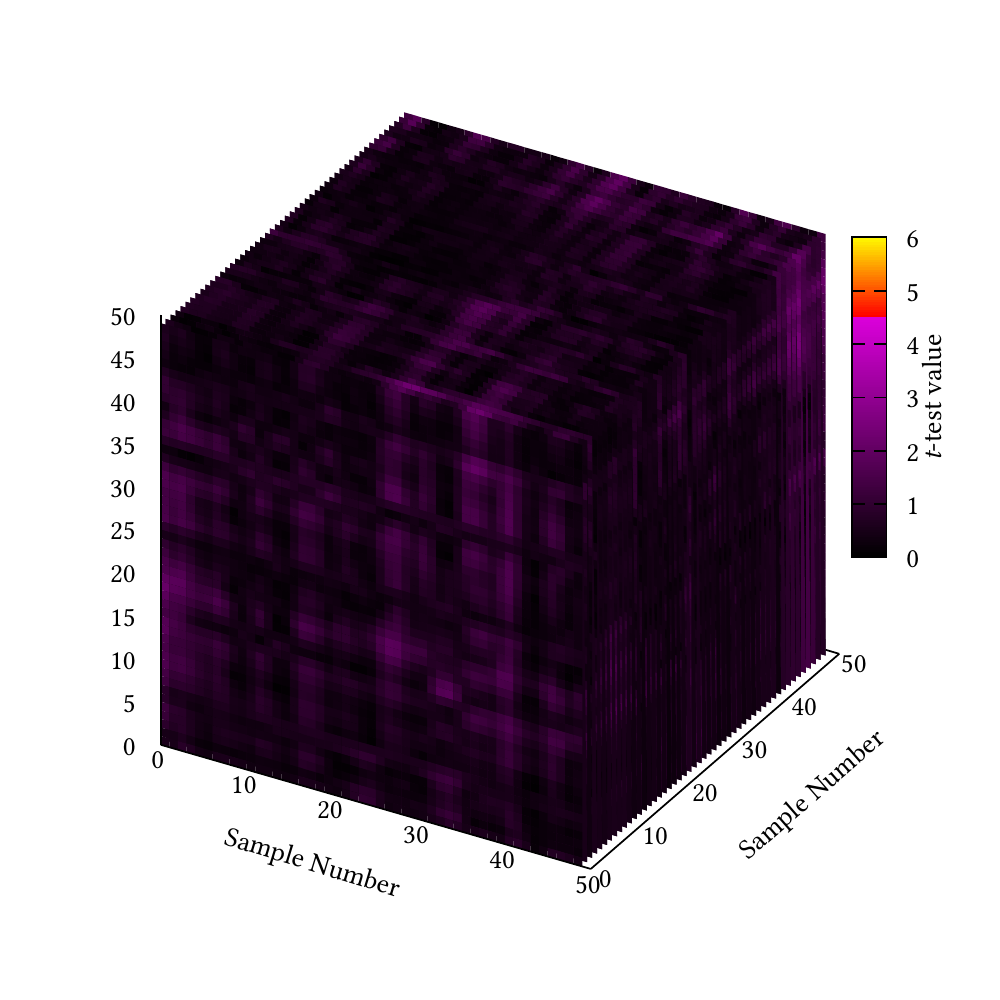}
  \end{center}
		\caption{After applying code fixes. $t$-value peak of 3.31 at (19,0,6)\label{f:synthetic4_after}}
	\end{subfigure}
	\caption{Evaluating a toy example. }
	\label{f:synthetic4}
\end{figure}

In \cref{f:synthetic_before} we see the results of bivariate leakage analysis on two million traces 
collected using our experimental setup.
The figure is a heatmap, where the X and Y axes indicate the samples that are combines to create the artificial bivariate sample.
The colour of each combined sample indicates the magnitude of the fixed vs.\ 
random \ttest analysis for the combined sample. The figure is symmetric across 
its main diagonal.

Examining the figure we find that there are two regions that show a \ttest value above our threshold of 4.5.
These occur at the combinations of samples around 50, which corresponds to \cref{line:synth1} of \cref{l:example}, and sample 200, which corresponds to \cref{line:synth3}.
Running \horosita also shows that the combination of  \cref{line:synth1} and \cref{line:synth3} leaks. 
Root-cause analysis shows that \cref{line:synth2,line:synth3} interact both through the processor pipeline and through the memory bus.

\looseness=-1 
To fix the leakage, \horosita first inserts a \texttt{mov r7, r7} instruction between \cref{line:synth2} and \cref{line:synth3}, and repeats the analysis to check that the leakage has been eliminated.
Finding that there is still leakage through the memory bus, \horosita further 
adds a combination of \texttt{push} and \texttt{pop} instructions, producing 
the code in \cref{l:fixedtoy}. \horosita required 200,000 emulated traces to 
apply fixes for this implementation.
Running the bivariate analysis on the code shows no evidence of second-order leakage, as shown in \cref{f:synthetic_after}.

\begin{lstlisting}[
  numbers=left,
  numberstyle=\scriptsize,
  keepspaces,
  basicstyle=\footnotesize\ttfamily,
  xleftmargin=2em,
  caption=Fixed Toy Example,
  label=l:fixedtoy,
  escapechar=|]
  ; nop padding
mov r7, r7
ldrb r4, [r1]
push {r7}    
pop {r7}   
  ; nop padding 
ldrb r5, [r2] 
mov r7, r7
push {r7}
pop{r7}
ldrb r6, [r3]
  ; nop padding
\end{lstlisting}

We 
extended the second order Boolean masked implementation shown 
in~\cref{l:example} to the third order by 
introducing another share to it. The code for this implementation is shown 
in~\cref{l:example4}. Similar to the second order version, we intentionally design
the example to leak operand information from the last two \texttt{ldr} 
instructions. This implementation was fixed by \horosita with two million 
emulated traces. To detect leakage in the physical device we had to collect 30 
million traces. \cref{f:synthetic4_before} shows the detected leakage from a 
3rd order \ttest. A first order \ttest was run on the combined traces that were 
combined using \cref{eq:biv} with a window of 50 samples. 
In contrast, \cref{f:synthetic4_after} shows the results of the \ttest that was run on 30 
million side-channel traces taken from the physical experiment after applying \horosita{}'s 
fixes.

\begin{lstlisting}[
	numbers=left,
	numberstyle=\scriptsize,
	keepspaces,
	xleftmargin=2em,
  basicstyle=\footnotesize\ttfamily,
	caption=A Toy Example (third order),
	label=l:example4,
	escapechar=|]
	ldr r3, [r1,#0] 
	push {r7}     
	pop {r7}
	; nop padding
	ldr r4, [r1,#4]
	push {r7}     
	pop {r7}
	; nop padding
	ldr r5, [r1,#16]
	ldr r6, [r1,#20]
	; nop padding
\end{lstlisting}

\parhead{Comparing Emulated and Real Traces:}
To better understand the relationship between emulated and real traces,
we compared the leakage observed in the traces in terms of signal-to-noise 
ratio (SNR). 
For this experiment we used 20,000 random input traces coming from the 
emulation and the real 
experiments using the code segment shown in~\cref{l:example}; we chose this 
number because it is sufficient to find leakage in the emulated traces 
using TVLA. 
We computed SNR for the leaking values that need to be combined for bivariate 
analysis: hamming weight (HW) of 4 bytes of $r1$ and HW of 4 bytes of $r2 
\oplus r3$. 
For the real experiments these values were between $0.041$ and $0.063$ for the 
bytes of $r1$ and between $0.012$ and $0.014$ the bytes of $r2 \oplus r3$. 
We could not compute the SNR directly for the emulated traces since the 
emulation is deterministic 
and therefore, noise-free. 
We added a sufficient amount of noise to generate similar SNR to the real 
experiments. 
We used Gaussian Noise with means $0$ and standard deviation of $0.25$\% of the 
signal amplitude 
for the bytes of $r1$ and $0.1$\% for the bytes of $r2 \oplus r3$. 
We do not know from where this leakage difference is exactly coming from, but 
we suspect that we simply 
found a slight difference between the emulated and the real measurements.

We conclude that if we introduce between $0.1$\% and $0.25$\% ratio of noise to 
the emulated traces then we obtain a similar SNR to the real traces.
Moreover, we can use $25$ times less traces than in the real experiment to 
detect leakage using 
TVLA, since we can detect leakage using emulation with 20,000 traces and we 
need 500,000 in the 
real experiments.

\subsection{Evaluated Cryptographic Implementations} \label{ss:prep}

We now turn our attention to more realistic examples. 
Before performing the evaluation we use \rosita to detect and eliminate any first-order leakage 
from the code. 
We further perform a first-order fixed vs.\ random TVLA with 2,000,000 traces on the real hardware 
to verify that no first-order leakage is detected.
For the evaluation, we use \horosita to detect and correct second-order leakage for 500,000 
simulated traces.
We then collect 2,000,000 power traces from each of  the original and the fixed software,
and perform bivariate second-order analysis to identify any leakage.
We evaluate two cryptographic implementations and one cryptographic primitive, 
which we describe below.

\parhead{Xoodoo}\label{s:xoodoo}
Xoodoo was proposed by \citet{DaemenHAK18} and a reference 
implementation is available from \citet{bertoniextended}. 
We converted this code to a 
three-share implementation 
based on the Threshold Implementation (TI) approach~\cite{DBLP:conf/icics/NikovaRR06}.
TI schemes were proposed 
to prevent the leakage from ``glitches'' that can occur in hardware implementations. 
The concept is accomplishing the goal of masking through a number of shares with some additional properties. Specifically, the non-completeness property of TI enforces that no operation should involve more than two shares.

Xoodoo's state is 48 bytes in length. The state is divided into three equal 
blocks called \emph{planes}, each consisting  of four 32-bit words. $x_{i,j}$ denotes the $j$\textsuperscript{th} 32-bit word of the
$i$\textsuperscript{th} plane of share $x$, where $x~\in~\{a,b,c\}$. 
\cref{l:xoodoo3} shows the algorithm segment that we analyse,
which forms part of the start of the Xoodoo $\chi$ function. 
Our initial C implementation demonstrated first-order leakage caused by the optimiser merging shares. 
We therefore manually implemented the code in assembly, ensuring that shares are not merged.

\begin{lstlisting}[aboveskip=-0.8 \baselineskip, belowskip=-0.8 \baselineskip,float,label=l:xoodoo3,numberstyle=\scriptsize,caption=Xoodoo code segment under test, basicstyle=\footnotesize\ttfamily,
escapeinside={(*}{*)}]
(* $$a_{0,0} = a_{0,0} \oplus (\neg a_{1,0} \land a_{2,0}) \oplus (a_{1,0} \land b_{2,0}) 
\oplus (b_{1,0} \land a_{2,0})$$%
$$b_{0,0} = b_{0,0} \oplus (\neg b_{1,0} \land b_{2,0}) \oplus (b_{1,0} \land c_{2,0}) 
\oplus (c_{1,0} \land b_{2,0})$$%
$$c_{0,0} = b_{0,0} \oplus (\neg c_{1,0} \land c_{2,0}) \oplus (c_{1,0} \land a_{2,0}) 
\oplus (a_{1,0} \land c_{2,0})$$ *)
\end{lstlisting}

\parhead{\Present}
\Present is a block cipher based on a substitution permutation network, which was proposed by 
Bogdanov et al.\ in~\cite{BogdanovKLPPRSV07}.
It has a block size of 64-bit and the key can be 80 or 128 bits long.
The non-linear layer is based on a single 4-bit S-box facilitating lightweight hardware implementations.

We implemented \present with
side-channel protection in software based on TI with three shares, as described 
by~\citet[Alg.\ 3.2]{SasdrichB018}. 
Thus, at least in theory, the implementation should not leak in the first order.
We used the code 
shown in \cref{l:present3} that 
implements a part of the \present S-box, involving 3 shares $x^1,x^2,x^3$ and the lookup table $T$.
The table is an 8-bit to 4-bit lookup table where the inputs are two 4-bit nibbles.
Each table lookup used to compute $t^i$ is repeated 16 times to cover the complete 64-bit shares.

\begin{lstlisting}[aboveskip=-0.8 \baselineskip, belowskip=-0.8 \baselineskip, float,label=l:present3,numberstyle=\scriptsize,  keepspaces, basicstyle=\footnotesize\ttfamily,
caption=\Present code segment under 
test,escapeinside={(*}{*)}]
(*$$t^3 = \text{T}(x^1,x^2)$$
$$t^2 = \text{T}(x^3,x^1)$$
$$t^1 = \text{T}(x^2,x^3)$$
*)
\end{lstlisting}

Observe that threshold implementations with three shares provides provable first-order security, but only limited protection against the second-order attacks~\cite{DBLP:conf/icics/NikovaRR06}. 
Therefore, we can expect that diminished second-order leakage may occur for both Xoodoo and \present implementations.

\parhead{Second-order Boolean-to-arithmetic masking} 
Boolean-to-arithmetic masking~\cite{GoubinCHES2001} is a cryptographic building block that converts a Boolean mask to an arithmetic mask. 
It is often used in side-channel resistant implementations  of cryptographic algorithms that mix Boolean and arithmetic operations (for example, ChaCha~\cite{Bernstein2008}). 
We implement and evaluate the second-order Boolean-to-arithmetic masking of~\citet[Alg.\ 2]{HutterJCEN2019}. 

In our evaluation this procedure takes as input boolean shares $x' = x \oplus 
r_1 \oplus r_2$, where $x$, $r_1$ and $r_2$ are random in 
$\mathbb{Z}_{2^{32}}$. For side-channel protection, the procedure uses
three additional masks $\gamma_1$, $\gamma_2$, and $\alpha$ also random in 
$\mathbb{Z}_{2^{32}}$. It computes $x'' = x + s_1 + s_2$, where $x''$, $s_1$, 
and $s_2$ 
are the output arithmetic shares. 
This implementation is proven to be second-order secure in~\cite{HutterJCEN2019} and therefore, we do not expect to see leakage in an implementation protected with \horosita. 

\subsection{Emulation results} \label{s:emuresults}
\begin{figure}[htb]
	\begin{subfigure}{0.25\textwidth}	
		\begin{center}
		\includegraphics[scale=0.45]{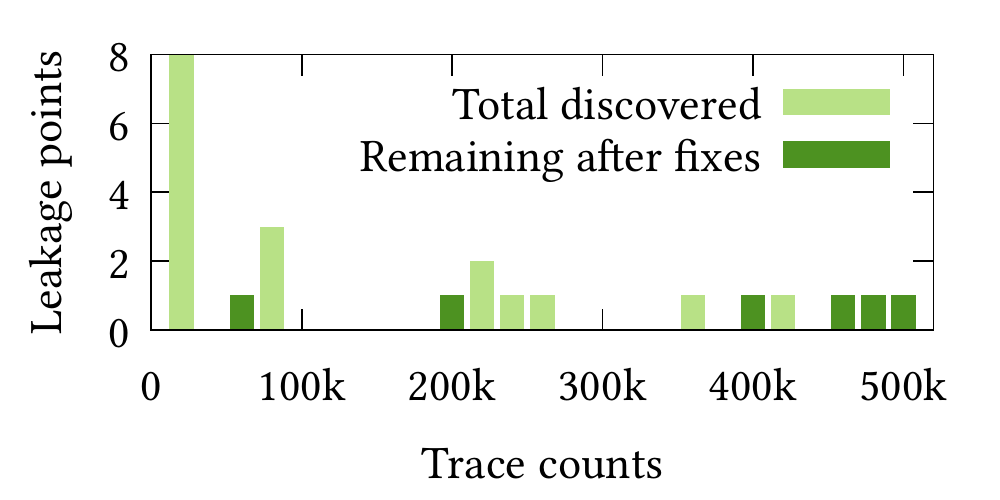}
		\end{center}
		\vspace{-.4em}
		\caption{Xoodoo}
		\label{f:xoodoo_discvsfix_prot}
	\end{subfigure}%
	\begin{subfigure}{0.25\textwidth}
		\begin{center}
		\includegraphics[scale=0.45]{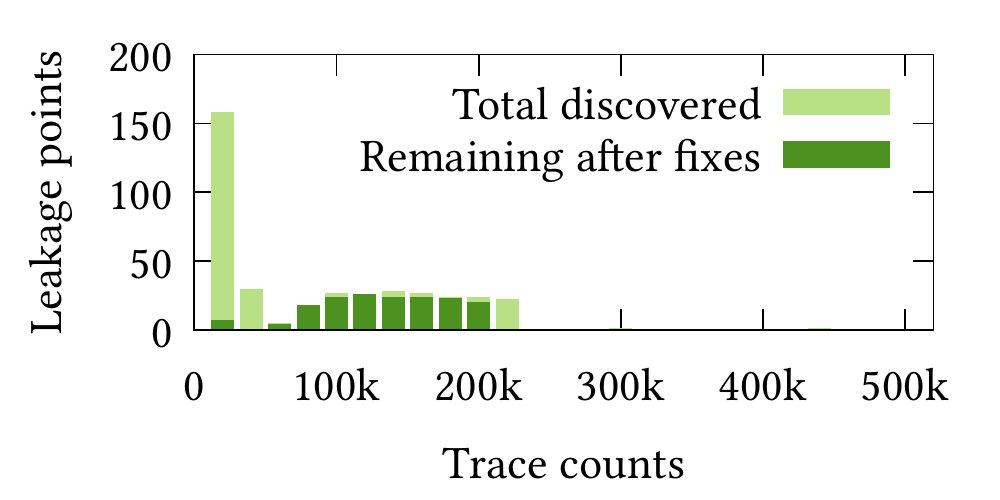}
		\end{center}
		\vspace{-.4em}
		\caption{\Present}
		\label{f:present_discvsfix_prot}
	\end{subfigure}

	\vspace{-.3em}
	\begin{subfigure}{0.5\textwidth}
		\begin{center}
			\includegraphics[scale=0.45]{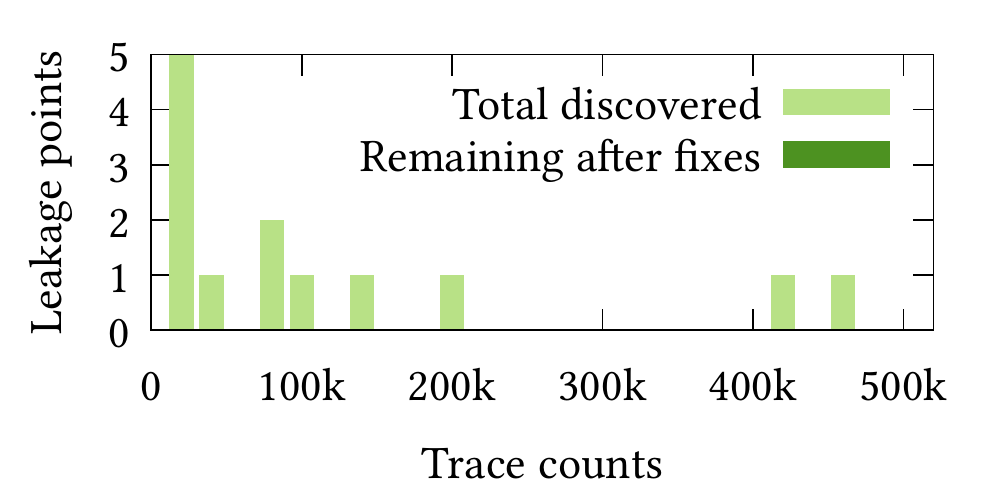}
		\end{center}
		\vspace{-.4em}
		\caption{Boolean-to-arithmetic}
		\label{f:bool2arith_discvsfix_prot}
	\end{subfigure}
	\vspace{-.5em}

	\caption{Discovered and remaining leakage points when fixed code from 
	previous iteration is used as input to the next iteration of \horosita}
	\label{f:discvsfix_prot}
	\vspace{-.5em}
\end{figure}

We used \horosita to fix the leakages that were discovered in the code segments 
introduced 
above in \cref{ss:prep}. Specifically, we focus on leakage discovered by a 
bivariate fixed 
vs.\ random \ttest.

To analyse the relationship between the number of traces and leakage discovery, 
we ran \horosita on 
the unprotected ciphers, varying the number of traces from 20,000 to 500,000 at 
steps of 20,000. 
In each iteration we used the output of the previous iteration as the input.
Each 
iteration performed emulation and root-cause detection. The emulation results 
are shown in~\cref{f:discvsfix_prot}. This proves to be more efficient than running 
\horosita a single time with a large number emulation traces. The reason for 
the efficiency of the iterations based method with gradually increasing trace 
counts is that leakage is fixed as it is detected so that large numbers of 
traces are not required for fixing all the leakage points 
that are detected. 
\cref{t:rositadurations} shows the emulation and root cause detection time when fixed code is used from the previous iteration.
\cref{t:overhead} shows the performance overhead of the code fixes.

\begin{table}
  {\small
	\begin{tabular}{lrr}
		\toprule
		\multicolumn{1}{l}{\bfseries \small Implementation} & 
		\multicolumn{1}{r}{\bfseries \small Emulation time} &
		\multicolumn{1}{r}{\bfseries \small Root Cause Det. time} \\
		\midrule
		Xoodoo & 1:35:41 & 3:12 \\
		\Present & 1:55:19 & 24:46 \\
		Boolean to arithmetic & 1:08:19 & 1:07 \\
		\bottomrule
		
	\end{tabular}
  }
	\caption{Time taken for emulation and root-cause detection}
	\label{t:rositadurations}
\end{table}

\begin{table}
  {\small
	\begin{tabular}{lrrr}
		\toprule
		\multicolumn{1}{l}{\bfseries \small Implementation} & 
		\multicolumn{1}{r}{\bfseries \small Unprotected} &
		\multicolumn{1}{r}{\bfseries \small Protected} & 
		\multicolumn{1}{r}{\bfseries \small Increase} \\
		&
		\multicolumn{1}{r}{\bfseries \small  size (cycles)} &
		\multicolumn{1}{r}{\bfseries \small  size (cycles)} &
		\multicolumn{1}{r}{}\\
		\midrule
		Xoodoo & 56 & 76 & 36\%\\
		\Present & 114 & 330 & 189\%\\
		Boolean to arithmetic & 75 & 97 & 29\% \\
		\bottomrule
		
	\end{tabular}
  }
	\caption{Performance overhead of fixes}
	
	\label{t:overhead}
\end{table}

We observe that after emulation 500,000 traces for the fixed vs.\ random 
\ttest{}, there was only one remaining leakage in the Xoodoo masked 
implementation. \Present and Boolean-to-arithmetic implementations did not have 
any remaining leakage points. However, when running the physical experiments we 
observed that the remaining leakage in Xoodoo was not significant.

\begin{lstlisting}[numbers=left,numberstyle=\scriptsize,keepspaces,xleftmargin=2em,caption=Leaky 
code segment of fixed \present, basicstyle=\footnotesize\ttfamily, label=l:presentleak,escapechar=|]
ldrb	r2, [r4, #16]
lsls	r1, r1, #4
adds	r1, r3, r1
ldrb	r0, [r1, r2] |\label{l:line_leak_ldr0}|
\end{lstlisting}

\begin{table}
  {\small
  \begin{tabular}{lrr}
	  \toprule
		\bfseries \small Trace set & 
	          \multicolumn{1}{r}{\bfseries \small Samples} &
	          \multicolumn{1}{c}{\bfseries \small Wall Clock Time} \\
	  \midrule
		Xoodoo unprotected & 1000 & 4:51\\
		Xoodoo protected & 1400 & 33:50  \\
		Present unprotected & 1400 & 28:31\\
		Present protected & 3500 & 7:02:00\\
		Boolean to arithmetic unprotected & 1000 & 4:18\\
		Boolean to arithmetic protected & 1200 & 8:51\\
	  \bottomrule
	\end{tabular}
  }
	\caption{Bivariate analysis time}
	\label{t:perf}
\end{table}

\subsection{Physical experiment results}

\begin{figure*}[htb]
  \begin{subfigure}[t]{.28\linewidth}
    \begin{center}
      \includegraphics[scale=0.5]{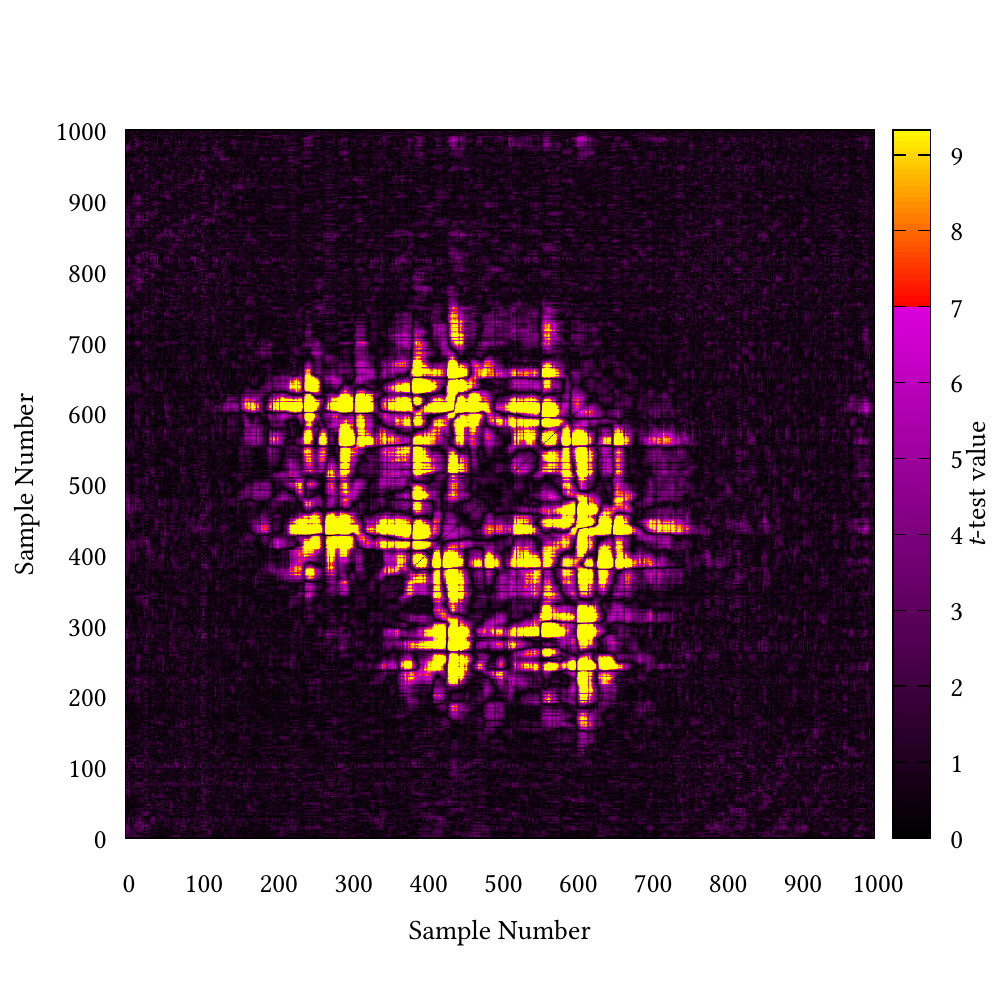}
    \end{center}
    \vspace*{-0.5cm}\caption{Xoodoo: before applying code fixes, $t$-value peak: 70.32}
    \label{f:xoodoo_before}
  \end{subfigure}\hspace{.05\linewidth}%
  \begin{subfigure}[t]{.28\linewidth}
    \begin{center}
      \includegraphics[scale=0.5]{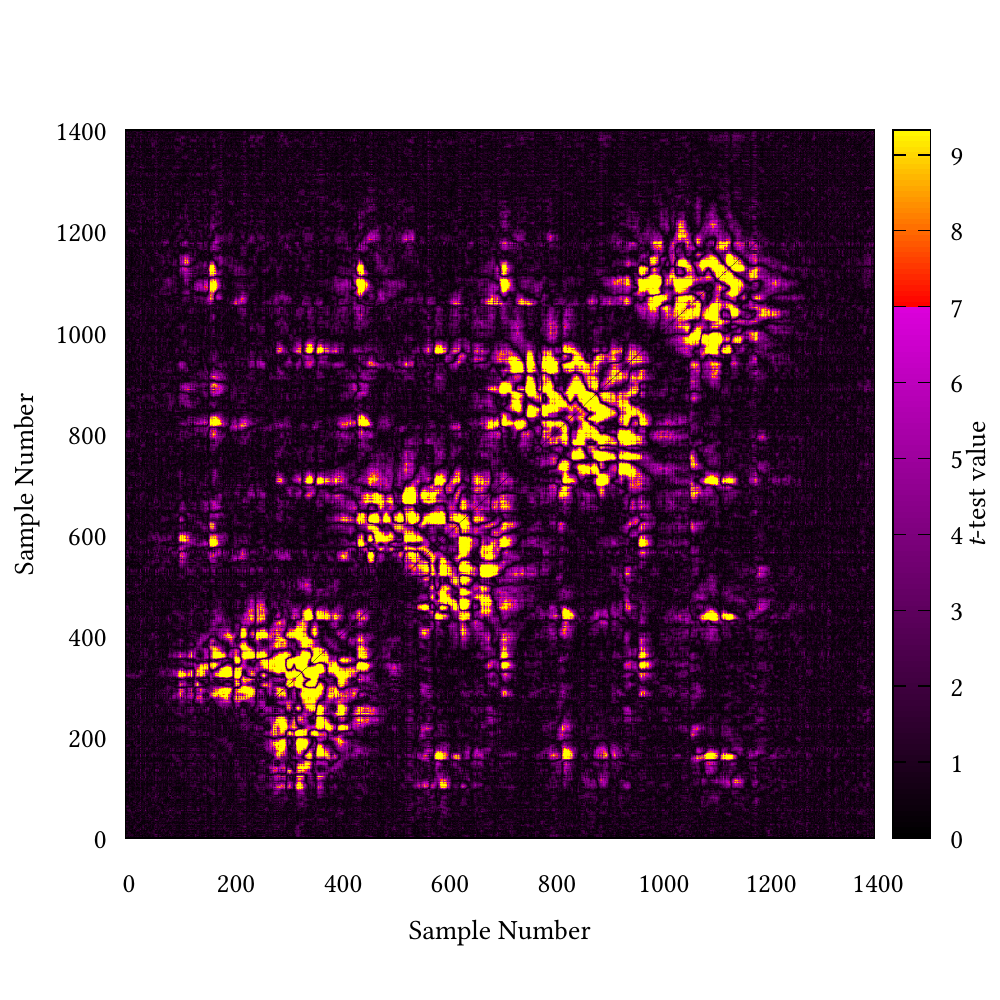}
    \end{center}
    \vspace*{-0.5cm}\caption{\Present: before applying code fixes, $t$-value peak: 55.13 }
    \label{f:present_before}
  \end{subfigure}\hspace{.05\linewidth}%
  \begin{subfigure}[t]{.28\linewidth}
    \begin{center}
      \includegraphics[scale=0.5]{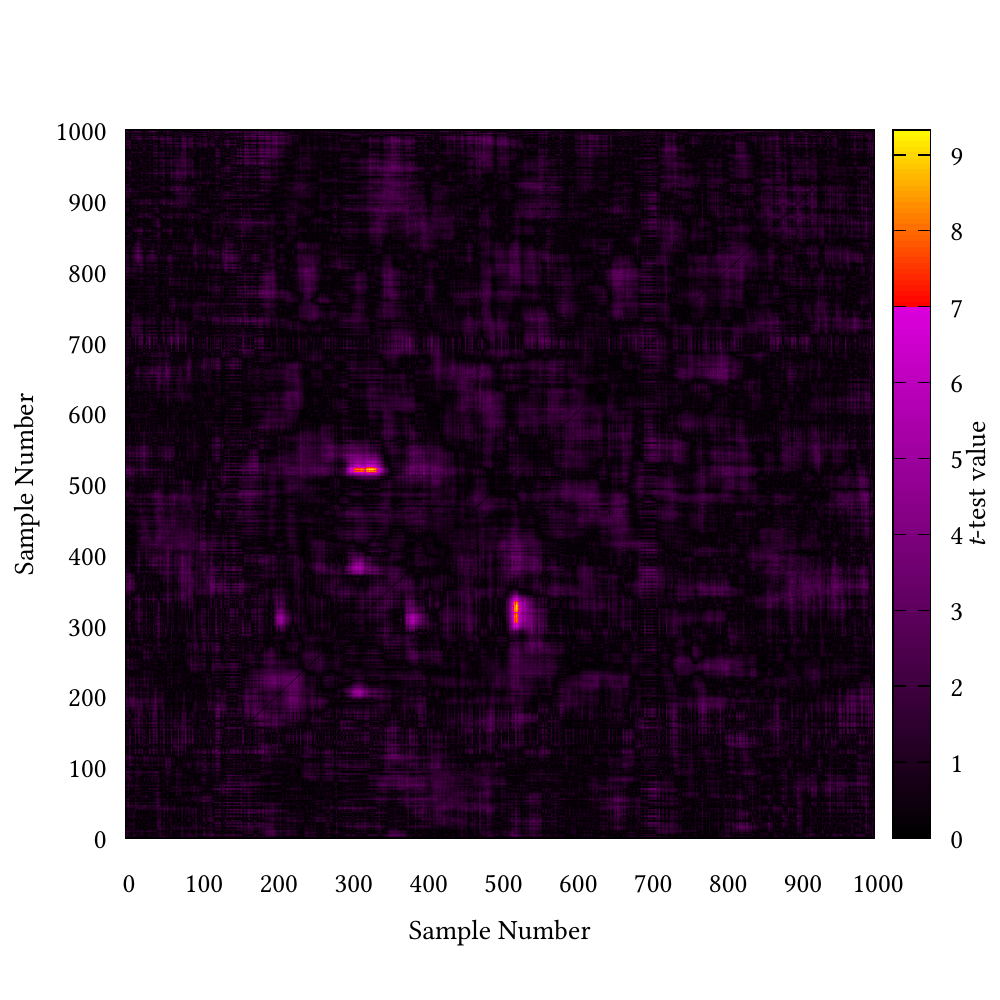}
    \end{center}
      \vspace*{-0.5cm}\caption{B-to-A: before applying code fixes, $t$-value peak: 9.13 }
      \label{f:bool2arith_before}
  \end{subfigure}

  \vspace{-.5em}
  \begin{subfigure}[t]{.28\linewidth}
    \begin{center}
      \includegraphics[scale=0.5]{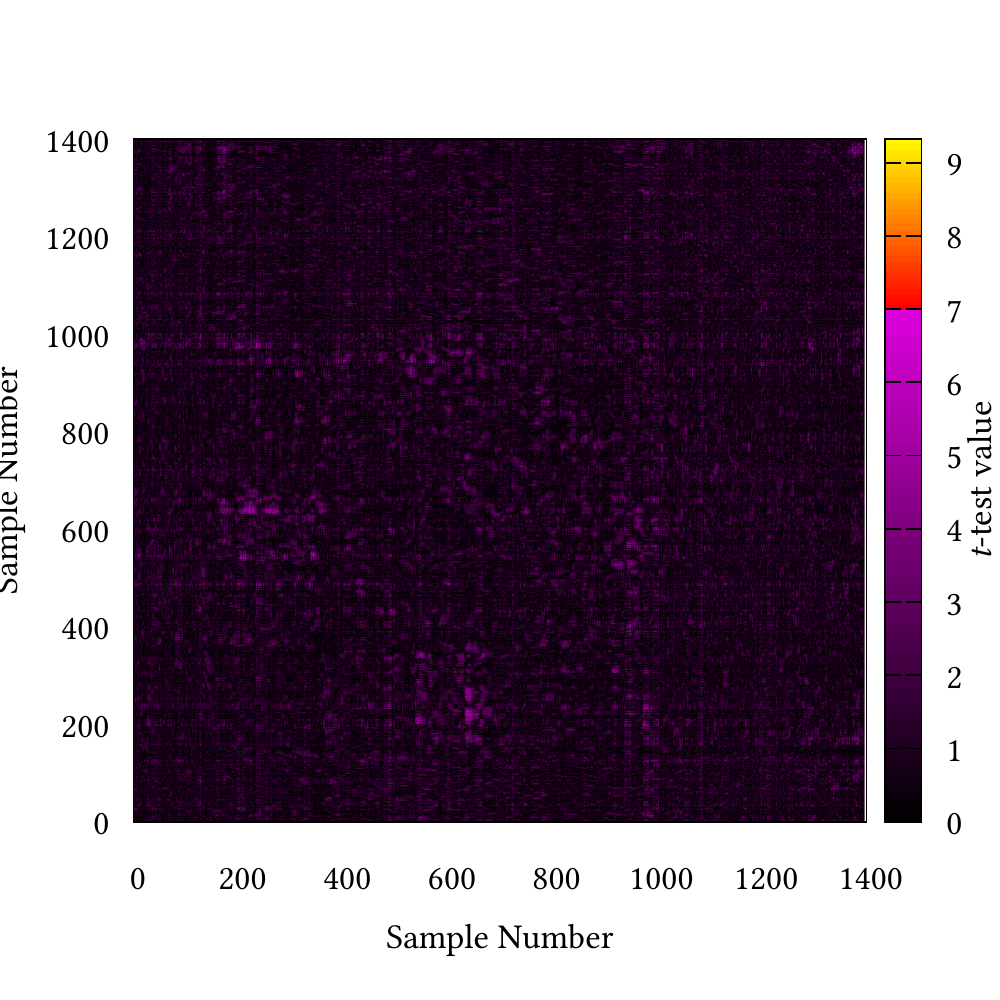}
    \end{center}
    \vspace*{-0.5cm}\caption{Xoodoo: after applying code fixes, $t$-value peak: 6.44 }
    \label{f:xoodoo_after}
  \end{subfigure}\hspace{.05\linewidth}%
  \begin{subfigure}[t]{.28\linewidth}
    \begin{center}
      \includegraphics[scale=0.5]{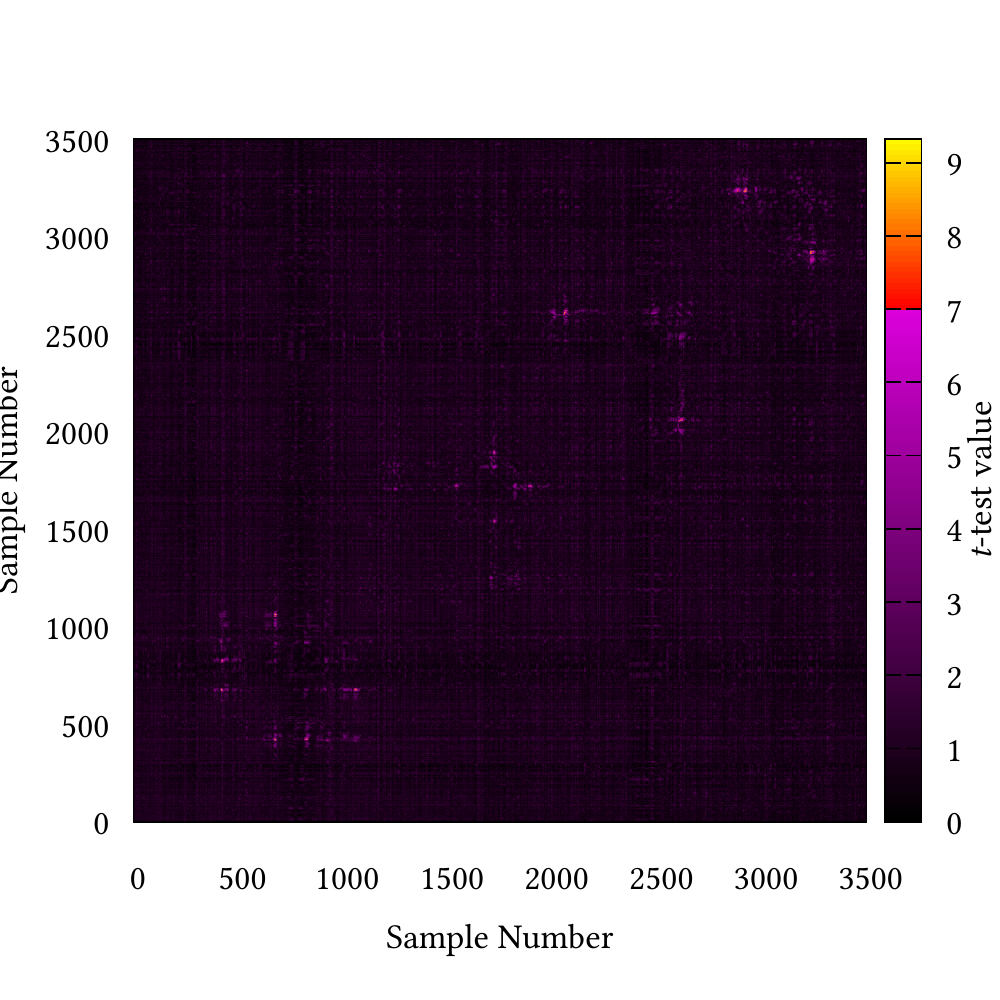}
    \end{center}
    \vspace*{-0.5cm}\caption{\Present: after applying code fixes, $t$-value peak: 12.38 }
    \label{f:present_after}
  \end{subfigure}\hspace{.05\linewidth}%
    \begin{subfigure}[t]{.28\linewidth}
    \begin{center}
      \includegraphics[scale=0.5]{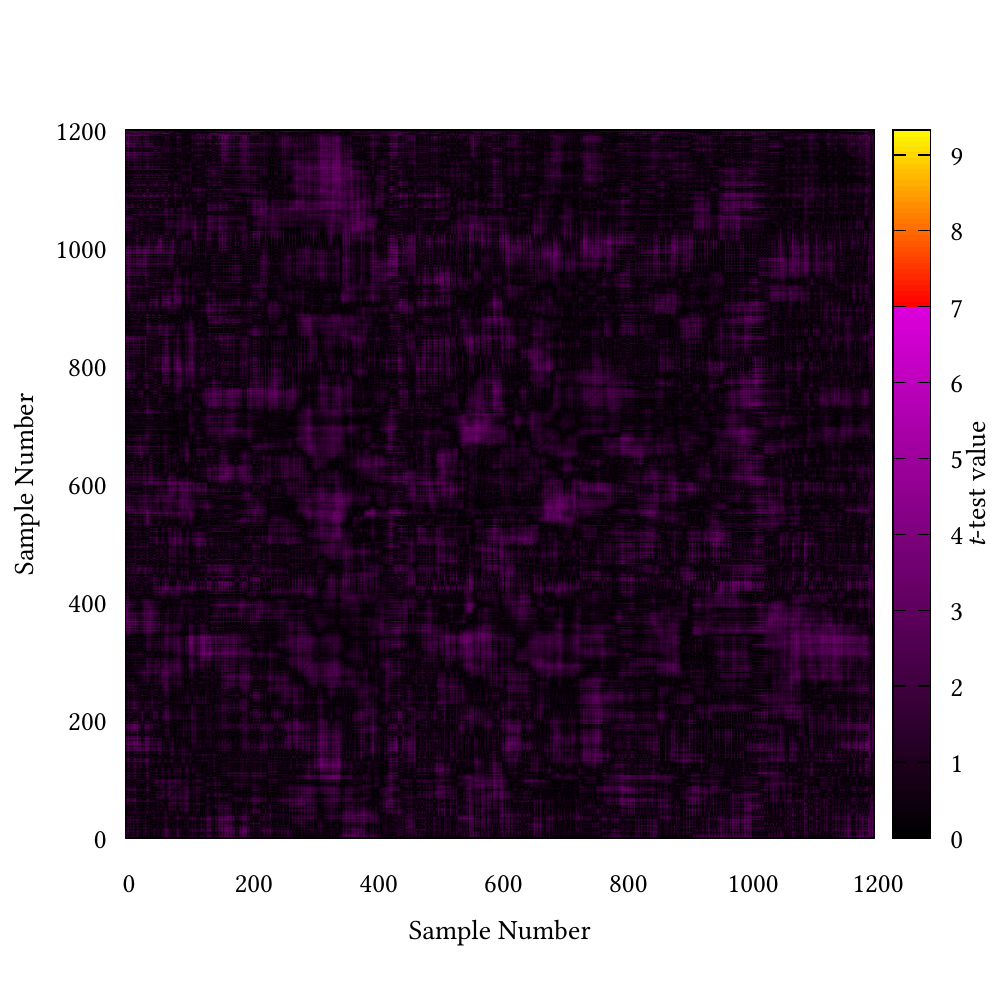}
    \end{center}
    \vspace*{-0.5cm}\caption{B-to-A: after applying code fixes, $t$-value peak: 3.91 }
    \label{f:bool2arith_after}
  \end{subfigure}
  \vspace{-.5em}

  \caption{Evaluation of three cryptographic primitives (B-to-A stands for Bollean to arithmetic)}
	\label{f:three_crypt}
\end{figure*}

\cref{f:three_crypt} compares the \ttest values of side-channel traces for the three ciphers before and after \horosita, as measured on the physical device.
The top row (\cref{f:xoodoo_before,f:present_before,f:bool2arith_before}) show the leakage of the original implementations, whereas the bottom row (\cref{f:xoodoo_after,f:present_after,f:bool2arith_after}) shows the leakage after applying \horosita.
The three implementations were protected using 500,000 emulated traces. 
Collecting the traces took around 8 hours for \present and for Boolean-to-arithmetic, and around 9:30 hours for Xoodoo, which requires significantly more mask bytes, slowing down the communication with the PC.

As the figures show, for Xoodoo and the Boolean-to-arithmetic masking conversion, \horosita completely eliminates leakage. 
However, for \present some leakage is not fixed.
Further analysis shows that this leakage is caused by interactions through the address bus.
\cref{l:presentleak} 
shows the first leaky segment of the code corresponding to
samples (700, 440) in \cref{f:present_after}. We confirmed this 
leakage through correlation based testing against actual share values and their 
combinations. The registers 
used for addressing in the \texttt{ldrb} instruction at \cref{l:line_leak_ldr0} 
carry one share each. Our investigation showed that sample 440 originates from 
this point. Additionally, the missing share is provided by the instruction that 
corresponds to sample 700. Both points show high correlation to the share 
values. Therefore, this leakage becomes observable as second-order 
leakage. We confirmed this leakage pattern by reproducing the same 
effect in a separate fixed vs.\ random experiment which has only two shares 
that is used in an \texttt{ldrb} instruction for addresses. It 
showed significant first order leakage at 200,000 traces.

Because our tooling 
does not detect address leakage, this code cannot be currently corrected. Moreover, we suspect that that the leakage might be present here due to the used threshold implementation algorithm and therefore, solving it is out of the scope of this work. 

\subsection{Tools for Leakage Analysis}
We now present the performance of our second-order analysis tools.
We run the tools on a desktop computer, featuring an Intel Core i9-10900K CPU 
and 32\,GB of memory. We spawn 10 threads and perform bivariate analysis of 
four cryptographic implementations.
For each implementation we use our measurement setup to collect 2M traces from 
the real experiments, which we analyse to draw the heatmaps shown 
in~\cref{f:three_crypt}.
The results are shown in \cref{t:perf}. The number of threads used can be 
changed to fit the underlying hardware, the thread count is dependent on the 
equal sized splits that are done along 
the sample axis. It is given by $S(S+1) \div 2$ where $S$ is the number of equal sized splits.
For our runs, $S$ was set at 4. 
Without parallelisation the run time will be 8 times slower if 
run in a single thread as 4 out of 10 of the threads do half of the work.

\section{Conclusions}
\label{sec:conclusions}

Since the introduction of side-channel attacks, implementation security of embedded devices has been under the immense 
scrutiny and constant threat of being exploited. Even with theoretically sound measures such as masking, the devices tend to exhibit some leakages in practice due to unintended interactions in hardware. 
Mostly manual evaluation involving a tedious decision process and applying fixes to such ``leaky'' implementations have since been adopted. Some automatic countermeasures have also been developed, but 
all of them target univariate leakage.  

\looseness=-1 
In this work, we set out to automate the detection and application of fixes for high order secured implementations through 
multivariate analysis. We have demonstrated that it is possible to fix 
almost all detected leakage for three second-order masked 
implementations 
using our 
root cause analysis. 
Furthermore, we have shown practically that our methodology also is applicable for the third order analysis.
It is a significant improvement over previous automatic countermeasure application methods due to its simplicity. 

\FloatBarrier

\ifAnon
\else
\section*{Acknowledgements}
We thank the anonymous reviewers for their insightful comments and recommendations.

This work was supported by
ARC Discovery Early Career Researcher Award number DE200101577; 
ARC Discovery Projects numbers DP200102364 and DP210102670; 
the Blavatnik ICRC at Tel-Aviv University; 
European Commission through the ERC Starting Grant 805031 (EPOQUE) of Peter Schwabe; 
and gifts from Facebook, Google 
and Intel.

Parts of this work were carried out while Yuval Yarom was affiliated with CSIRO's Data61.
\fi

\bibliographystyle{ACM-Reference-Format}
\bibliography{horosita}


\begin{thebibliography}{68}


\ifx \showCODEN    \undefined \def \showCODEN     #1{\unskip}     \fi
\ifx \showDOI      \undefined \def \showDOI       #1{#1}\fi
\ifx \showISBNx    \undefined \def \showISBNx     #1{\unskip}     \fi
\ifx \showISBNxiii \undefined \def \showISBNxiii  #1{\unskip}     \fi
\ifx \showISSN     \undefined \def \showISSN      #1{\unskip}     \fi
\ifx \showLCCN     \undefined \def \showLCCN      #1{\unskip}     \fi
\ifx \shownote     \undefined \def \shownote      #1{#1}          \fi
\ifx \showarticletitle \undefined \def \showarticletitle #1{#1}   \fi
\ifx \showURL      \undefined \def \showURL       {\relax}        \fi
\providecommand\bibfield[2]{#2}
\providecommand\bibinfo[2]{#2}
\providecommand\natexlab[1]{#1}
\providecommand\showeprint[2][]{arXiv:#2}

\bibitem[\protect\citeauthoryear{Aumasson, Henzen, Meier, and Phan}{Aumasson
  et~al\mbox{.}}{2009}]%
        {Aumasson2009SHA3PB}
\bibfield{author}{\bibinfo{person}{Jean-Philippe Aumasson}, \bibinfo{person}{L.
  Henzen}, \bibinfo{person}{W. Meier}, {and} \bibinfo{person}{R. Phan}.}
  \bibinfo{year}{2009}\natexlab{}.
\newblock \bibinfo{title}{SHA-3 proposal BLAKE}.
\newblock
\newblock


\bibitem[\protect\citeauthoryear{Aysu, Tobah, Tiwari, Gerstlauer, and
  Orshansky}{Aysu et~al\mbox{.}}{2018}]%
        {AysuTTGO18}
\bibfield{author}{\bibinfo{person}{Aydin Aysu}, \bibinfo{person}{Youssef
  Tobah}, \bibinfo{person}{Mohit Tiwari}, \bibinfo{person}{Andreas Gerstlauer},
  {and} \bibinfo{person}{Michael Orshansky}.} \bibinfo{year}{2018}\natexlab{}.
\newblock \showarticletitle{Horizontal side-channel vulnerabilities of
  post-quantum key exchange protocols}. In \bibinfo{booktitle}{\emph{{HOST}}}.
  \bibinfo{pages}{81--88}.
\newblock


\bibitem[\protect\citeauthoryear{Balasch, Gierlichs, Grosso, Reparaz, and
  Standaert}{Balasch et~al\mbox{.}}{2015}]%
        {BalaschGGRS14}
\bibfield{author}{\bibinfo{person}{Josep Balasch}, \bibinfo{person}{Benedikt
  Gierlichs}, \bibinfo{person}{Vincent Grosso}, \bibinfo{person}{Oscar
  Reparaz}, {and} \bibinfo{person}{Fran{\c{c}}ois{-}Xavier Standaert}.}
  \bibinfo{year}{2015}\natexlab{}.
\newblock \showarticletitle{On the Cost of Lazy Engineering for Masked Software
  Implementations}. In \bibinfo{booktitle}{\emph{{CARDIS}}}.
  \bibinfo{pages}{64--81}.
\newblock


\bibitem[\protect\citeauthoryear{Batina, Bhasin, Jap, and Picek}{Batina
  et~al\mbox{.}}{2019}]%
        {BatinaBJP19}
\bibfield{author}{\bibinfo{person}{Lejla Batina}, \bibinfo{person}{Shivam
  Bhasin}, \bibinfo{person}{Dirmanto Jap}, {and} \bibinfo{person}{Stjepan
  Picek}.} \bibinfo{year}{2019}\natexlab{}.
\newblock \showarticletitle{{CSI} {NN:} Reverse Engineering of Neural Network
  Architectures Through Electromagnetic Side Channel}. In
  \bibinfo{booktitle}{\emph{{USENIX} Security Symposium}}.
  \bibinfo{pages}{515--532}.
\newblock


\bibitem[\protect\citeauthoryear{Bernstein}{Bernstein}{2008}]%
        {Bernstein2008}
\bibfield{author}{\bibinfo{person}{Daniel Bernstein}.}
  \bibinfo{year}{2008}\natexlab{}.
\newblock \bibinfo{title}{{ChaCha}, a variant of {Salsa20}}.
\newblock
\newblock


\bibitem[\protect\citeauthoryear{Bernstein}{Bernstein}{2005}]%
        {Bernstein2005}
\bibfield{author}{\bibinfo{person}{Daniel~J Bernstein}.}
  \bibinfo{year}{2005}\natexlab{}.
\newblock \bibinfo{title}{Cache-timing attacks on {AES}}.
\newblock
  \bibinfo{howpublished}{\url{https://cr.yp.to/antiforgery/cachetiming-20050414.pdf}}.
\newblock


\bibitem[\protect\citeauthoryear{Bertoni, Daemen, Hoffert, Peeters, Van~Assche,
  and Van~Keer}{Bertoni et~al\mbox{.}}{[n.\,d.]}]%
        {bertoniextended}
\bibfield{author}{\bibinfo{person}{Guido Bertoni}, \bibinfo{person}{Joan
  Daemen}, \bibinfo{person}{Seth Hoffert}, \bibinfo{person}{Micha{\"e}l
  Peeters}, \bibinfo{person}{Gilles Van~Assche}, {and} \bibinfo{person}{Ronny
  Van~Keer}.} \bibinfo{year}{[n.\,d.]}\natexlab{}.
\newblock \bibinfo{title}{The eXtended {Keccak} Code Package ({XKCP})}.
\newblock
\newblock
\urldef\tempurl%
\url{https://github.com/XKCP/XKCP}
\showURL{%
\tempurl}


\bibitem[\protect\citeauthoryear{Bertoni, Zaccaria, Breveglieri, Monchiero, and
  Palermo}{Bertoni et~al\mbox{.}}{2005}]%
        {BZB+05}
\bibfield{author}{\bibinfo{person}{Guido Bertoni}, \bibinfo{person}{Vittorio
  Zaccaria}, \bibinfo{person}{Luca Breveglieri}, \bibinfo{person}{Matteo
  Monchiero}, {and} \bibinfo{person}{Gianluca Palermo}.}
  \bibinfo{year}{2005}\natexlab{}.
\newblock \showarticletitle{{AES} Power Attack Based on Induced Cache Miss and
  Countermeasure}. In \bibinfo{booktitle}{\emph{ITCC}}.
  \bibinfo{pages}{586--591}.
\newblock


\bibitem[\protect\citeauthoryear{Bogdanov, Knudsen, Leander, Paar, Poschmann,
  Robshaw, Seurin, and Vikkelsoe}{Bogdanov et~al\mbox{.}}{2007}]%
        {BogdanovKLPPRSV07}
\bibfield{author}{\bibinfo{person}{Andrey Bogdanov}, \bibinfo{person}{Lars~R.
  Knudsen}, \bibinfo{person}{Gregor Leander}, \bibinfo{person}{Christof Paar},
  \bibinfo{person}{Axel Poschmann}, \bibinfo{person}{Matthew J.~B. Robshaw},
  \bibinfo{person}{Yannick Seurin}, {and} \bibinfo{person}{C. Vikkelsoe}.}
  \bibinfo{year}{2007}\natexlab{}.
\newblock \showarticletitle{{PRESENT:} An Ultra-Lightweight Block Cipher}. In
  \bibinfo{booktitle}{\emph{{CHES}}}. \bibinfo{pages}{450--466}.
\newblock


\bibitem[\protect\citeauthoryear{Brier, Clavier, and Olivier}{Brier
  et~al\mbox{.}}{2004}]%
        {BrierCO04}
\bibfield{author}{\bibinfo{person}{Eric Brier}, \bibinfo{person}{Christophe
  Clavier}, {and} \bibinfo{person}{Francis Olivier}.}
  \bibinfo{year}{2004}\natexlab{}.
\newblock \showarticletitle{Correlation Power Analysis with a Leakage Model}.
  In \bibinfo{booktitle}{\emph{{CHES}}}. \bibinfo{pages}{16--29}.
\newblock


\bibitem[\protect\citeauthoryear{Buhan, Batina, Yarom, and Schaumont}{Buhan
  et~al\mbox{.}}{2021}]%
        {BuhanBYS21}
\bibfield{author}{\bibinfo{person}{Ileana Buhan}, \bibinfo{person}{Lejla
  Batina}, \bibinfo{person}{Yuval Yarom}, {and} \bibinfo{person}{Patrick
  Schaumont}.} \bibinfo{year}{2021}\natexlab{}.
\newblock \showarticletitle{{SoK}: Design Tools for Side-Channel-Aware
  Implementations}.
\newblock \bibinfo{journal}{\emph{{IACR} Cryptol. ePrint Arch.}}
  \bibinfo{volume}{2021} (\bibinfo{year}{2021}), \bibinfo{pages}{497}.
\newblock


\bibitem[\protect\citeauthoryear{Chan, Golub, and Leveque}{Chan
  et~al\mbox{.}}{1983}]%
        {ChanGL1983}
\bibfield{author}{\bibinfo{person}{Tony~F. Chan}, \bibinfo{person}{Gene~H.
  Golub}, {and} \bibinfo{person}{Randall~J. Leveque}.}
  \bibinfo{year}{1983}\natexlab{}.
\newblock \showarticletitle{Algorithms for Computing the Sample Variance:
  Analysis and Recommendations}.
\newblock \bibinfo{journal}{\emph{The American Statistician}}
  \bibinfo{volume}{37}, \bibinfo{number}{3} (\bibinfo{year}{1983}),
  \bibinfo{pages}{242--247}.
\newblock
\urldef\tempurl%
\url{https://doi.org/10.1080/00031305.1983.10483115}
\showDOI{\tempurl}
\showeprint{https://www.tandfonline.com/doi/pdf/10.1080/00031305.1983.10483115}


\bibitem[\protect\citeauthoryear{Chari, Jutla, Rao, and Rohatgi}{Chari
  et~al\mbox{.}}{1999}]%
        {ChariJRR99}
\bibfield{author}{\bibinfo{person}{Suresh Chari}, \bibinfo{person}{Charanjit~S.
  Jutla}, \bibinfo{person}{Josyula~R. Rao}, {and} \bibinfo{person}{Pankaj
  Rohatgi}.} \bibinfo{year}{1999}\natexlab{}.
\newblock \showarticletitle{Towards Sound Approaches to Counteract
  Power-Analysis Attacks}. In \bibinfo{booktitle}{\emph{{CRYPTO}}}.
  \bibinfo{pages}{398--412}.
\newblock


\bibitem[\protect\citeauthoryear{Chen, Haider, and Schaumont}{Chen
  et~al\mbox{.}}{2009}]%
        {ChenHS09}
\bibfield{author}{\bibinfo{person}{Zhimin Chen}, \bibinfo{person}{Syed Haider},
  {and} \bibinfo{person}{Patrick Schaumont}.} \bibinfo{year}{2009}\natexlab{}.
\newblock \showarticletitle{Side-Channel Leakage in Masked Circuits Caused by
  Higher-Order Circuit Effects}. In \bibinfo{booktitle}{\emph{{ISA}}}.
  \bibinfo{pages}{327--336}.
\newblock


\bibitem[\protect\citeauthoryear{Chen and Zhou}{Chen and Zhou}{2006}]%
        {ChenZ06}
\bibfield{author}{\bibinfo{person}{Zhimin Chen} {and} \bibinfo{person}{Yujie
  Zhou}.} \bibinfo{year}{2006}\natexlab{}.
\newblock \showarticletitle{Dual-Rail Random Switching Logic: {A}
  Countermeasure to Reduce Side Channel Leakage}. In
  \bibinfo{booktitle}{\emph{{CHES}}}. \bibinfo{pages}{242--254}.
\newblock


\bibitem[\protect\citeauthoryear{Cnudde, Bilgin, Reparaz, Nikov, and
  Nikova}{Cnudde et~al\mbox{.}}{2015}]%
        {CnuddeBRNN15}
\bibfield{author}{\bibinfo{person}{Thomas~De Cnudde},
  \bibinfo{person}{Beg{\"{u}}l Bilgin}, \bibinfo{person}{Oscar Reparaz},
  \bibinfo{person}{Ventzislav Nikov}, {and} \bibinfo{person}{Svetla Nikova}.}
  \bibinfo{year}{2015}\natexlab{}.
\newblock \showarticletitle{Higher-Order Threshold Implementation of the {AES}
  S-Box}. In \bibinfo{booktitle}{\emph{{CARDIS}}}. \bibinfo{pages}{259--272}.
\newblock


\bibitem[\protect\citeauthoryear{Cnudde, Reparaz, Bilgin, Nikova, Nikov, and
  Rijmen}{Cnudde et~al\mbox{.}}{2016}]%
        {CnuddeRBNNR16}
\bibfield{author}{\bibinfo{person}{Thomas~De Cnudde}, \bibinfo{person}{Oscar
  Reparaz}, \bibinfo{person}{Beg{\"{u}}l Bilgin}, \bibinfo{person}{Svetla
  Nikova}, \bibinfo{person}{Ventzislav Nikov}, {and} \bibinfo{person}{Vincent
  Rijmen}.} \bibinfo{year}{2016}\natexlab{}.
\newblock \showarticletitle{Masking {AES} With d+1 Shares in Hardware}. In
  \bibinfo{booktitle}{\emph{TIS@CCS}}. \bibinfo{pages}{43}.
\newblock


\bibitem[\protect\citeauthoryear{Corre, Gro{\ss}sch{\"{a}}dl, and Dinu}{Corre
  et~al\mbox{.}}{2018}]%
        {CorreGD18}
\bibfield{author}{\bibinfo{person}{Yann~Le Corre}, \bibinfo{person}{Johann
  Gro{\ss}sch{\"{a}}dl}, {and} \bibinfo{person}{Daniel Dinu}.}
  \bibinfo{year}{2018}\natexlab{}.
\newblock \showarticletitle{Micro-architectural Power Simulator for Leakage
  Assessment of Cryptographic Software on {ARM} Cortex-M3 Processors}. In
  \bibinfo{booktitle}{\emph{{COSADE}}} \emph{(\bibinfo{series}{Lecture Notes in
  Computer Science}, Vol.~\bibinfo{volume}{10815})}.
  \bibinfo{publisher}{Springer}, \bibinfo{pages}{82--98}.
\newblock


\bibitem[\protect\citeauthoryear{Daemen, Hoffert, Assche, and Keer}{Daemen
  et~al\mbox{.}}{2018a}]%
        {DaemenHAK18}
\bibfield{author}{\bibinfo{person}{Joan Daemen}, \bibinfo{person}{Seth
  Hoffert}, \bibinfo{person}{Gilles~Van Assche}, {and}
  \bibinfo{person}{Ronny~Van Keer}.} \bibinfo{year}{2018}\natexlab{a}.
\newblock \showarticletitle{The design of Xoodoo and Xoofff}.
\newblock \bibinfo{journal}{\emph{{IACR} Trans. Symmetric Cryptol.}}
  \bibinfo{volume}{2018}, \bibinfo{number}{4} (\bibinfo{year}{2018}),
  \bibinfo{pages}{1--38}.
\newblock


\bibitem[\protect\citeauthoryear{Daemen, Hoffert, Assche, and Keer}{Daemen
  et~al\mbox{.}}{2018b}]%
        {DaemenHAK18C}
\bibfield{author}{\bibinfo{person}{Joan Daemen}, \bibinfo{person}{Seth
  Hoffert}, \bibinfo{person}{Gilles~Van Assche}, {and}
  \bibinfo{person}{Ronny~Van Keer}.} \bibinfo{year}{2018}\natexlab{b}.
\newblock \showarticletitle{Xoodoo cookbook}.
\newblock \bibinfo{journal}{\emph{{IACR} Cryptol. ePrint Arch.}}
  \bibinfo{volume}{2018} (\bibinfo{year}{2018}), \bibinfo{pages}{767}.
\newblock


\bibitem[\protect\citeauthoryear{den Hartog, Verschuren, de~Vink, de~Vos, and
  Wiersma}{den Hartog et~al\mbox{.}}{2003}]%
        {HartogVVVW03}
\bibfield{author}{\bibinfo{person}{Jerry den Hartog}, \bibinfo{person}{Jan
  Verschuren}, \bibinfo{person}{Erik~P. de Vink}, \bibinfo{person}{Jaap de
  Vos}, {and} \bibinfo{person}{W. Wiersma}.} \bibinfo{year}{2003}\natexlab{}.
\newblock \showarticletitle{{PINPAS:} {A} Tool for Power Analysis of
  Smartcards}. In \bibinfo{booktitle}{\emph{{IFIP SEC}}}.
  \bibinfo{pages}{453--457}.
\newblock


\bibitem[\protect\citeauthoryear{Ding, Zhang, Durvaux, Standaert, and Fei}{Ding
  et~al\mbox{.}}{2017}]%
        {DingZDSF17}
\bibfield{author}{\bibinfo{person}{A.~Adam Ding}, \bibinfo{person}{Liwei
  Zhang}, \bibinfo{person}{Fran{\c{c}}ois Durvaux},
  \bibinfo{person}{Fran{\c{c}}ois{-}Xavier Standaert}, {and}
  \bibinfo{person}{Yunsi Fei}.} \bibinfo{year}{2017}\natexlab{}.
\newblock \showarticletitle{Towards Sound and Optimal Leakage Detection
  Procedure}. In \bibinfo{booktitle}{\emph{{CARDIS}}}.
  \bibinfo{pages}{105--122}.
\newblock


\bibitem[\protect\citeauthoryear{Ferguson, Lucks, Schneier, Whiting, Bellare,
  Kohno, Callas, and Walker}{Ferguson et~al\mbox{.}}{2010}]%
        {Sklein2010}
\bibfield{author}{\bibinfo{person}{Niels Ferguson}, \bibinfo{person}{Stefan
  Lucks}, \bibinfo{person}{Bruce Schneier}, \bibinfo{person}{Doug Whiting},
  \bibinfo{person}{Mihir Bellare}, \bibinfo{person}{Tadayoshi Kohno},
  \bibinfo{person}{Jon Callas}, {and} \bibinfo{person}{Jesse Walker}.}
  \bibinfo{year}{2010}\natexlab{}.
\newblock \bibinfo{title}{The {Skein} Hash Function Family}.
\newblock
\newblock
\urldef\tempurl%
\url{https://www.schneier.com/academic/skein/}
\showURL{%
\tempurl}


\bibitem[\protect\citeauthoryear{Gandolfi, Mourtel, and Olivier}{Gandolfi
  et~al\mbox{.}}{2001}]%
        {GMO01}
\bibfield{author}{\bibinfo{person}{Karine Gandolfi},
  \bibinfo{person}{Christophe Mourtel}, {and} \bibinfo{person}{Francis
  Olivier}.} \bibinfo{year}{2001}\natexlab{}.
\newblock \showarticletitle{Electromagnetic Analysis: Concrete Results}. In
  \bibinfo{booktitle}{\emph{CHES}}. \bibinfo{pages}{251--261}.
\newblock


\bibitem[\protect\citeauthoryear{Gao, Marshall, Page, and Oswald}{Gao
  et~al\mbox{.}}{2020a}]%
        {GaoMPO20}
\bibfield{author}{\bibinfo{person}{Si Gao}, \bibinfo{person}{Ben Marshall},
  \bibinfo{person}{Dan Page}, {and} \bibinfo{person}{Elisabeth Oswald}.}
  \bibinfo{year}{2020}\natexlab{a}.
\newblock \showarticletitle{Share-slicing: Friend or Foe?}
\newblock \bibinfo{journal}{\emph{{IACR} Trans. Cryptogr. Hardw. Embed. Syst.}}
  \bibinfo{volume}{2020}, \bibinfo{number}{1} (\bibinfo{year}{2020}),
  \bibinfo{pages}{152--174}.
\newblock


\bibitem[\protect\citeauthoryear{Gao, Marshall, Page, and Pham}{Gao
  et~al\mbox{.}}{2020b}]%
        {GaoMPP20}
\bibfield{author}{\bibinfo{person}{Si Gao}, \bibinfo{person}{Ben Marshall},
  \bibinfo{person}{Dan Page}, {and} \bibinfo{person}{Thinh~Hung Pham}.}
  \bibinfo{year}{2020}\natexlab{b}.
\newblock \showarticletitle{{FENL:} an {ISE} to mitigate analogue
  micro-architectural leakage}.
\newblock \bibinfo{journal}{\emph{{IACR} Trans. Cryptogr. Hardw. Embed. Syst.}}
  \bibinfo{volume}{2020}, \bibinfo{number}{2} (\bibinfo{year}{2020}),
  \bibinfo{pages}{73--98}.
\newblock


\bibitem[\protect\citeauthoryear{Ge, Yarom, Cock, and Heiser}{Ge
  et~al\mbox{.}}{2018}]%
        {GeYCH18}
\bibfield{author}{\bibinfo{person}{Qian Ge}, \bibinfo{person}{Yuval Yarom},
  \bibinfo{person}{David Cock}, {and} \bibinfo{person}{Gernot Heiser}.}
  \bibinfo{year}{2018}\natexlab{}.
\newblock \showarticletitle{A survey of microarchitectural timing attacks and
  countermeasures on contemporary hardware}.
\newblock \bibinfo{journal}{\emph{J. Cryptogr. Eng.}} \bibinfo{volume}{8},
  \bibinfo{number}{1} (\bibinfo{year}{2018}), \bibinfo{pages}{1--27}.
\newblock


\bibitem[\protect\citeauthoryear{Genkin, Pachmanov, Pipman, Tromer, and
  Yarom}{Genkin et~al\mbox{.}}{2016}]%
        {GenkinPPTY16}
\bibfield{author}{\bibinfo{person}{Daniel Genkin}, \bibinfo{person}{Lev
  Pachmanov}, \bibinfo{person}{Itamar Pipman}, \bibinfo{person}{Eran Tromer},
  {and} \bibinfo{person}{Yuval Yarom}.} \bibinfo{year}{2016}\natexlab{}.
\newblock \showarticletitle{{ECDSA} Key Extraction from Mobile Devices via
  Nonintrusive Physical Side Channels}. In \bibinfo{booktitle}{\emph{{CCS}}}.
  \bibinfo{pages}{1626--1638}.
\newblock


\bibitem[\protect\citeauthoryear{Genkin, Shamir, and Tromer}{Genkin
  et~al\mbox{.}}{2014}]%
        {GenkinST14}
\bibfield{author}{\bibinfo{person}{Daniel Genkin}, \bibinfo{person}{Adi
  Shamir}, {and} \bibinfo{person}{Eran Tromer}.}
  \bibinfo{year}{2014}\natexlab{}.
\newblock \showarticletitle{{RSA} Key Extraction via Low-Bandwidth Acoustic
  Cryptanalysis}. In \bibinfo{booktitle}{\emph{{CRYPTO} {(1)}}}.
  \bibinfo{pages}{444--461}.
\newblock


\bibitem[\protect\citeauthoryear{Gigerl, Hadzic, Primas, Mangard, and
  Bloem}{Gigerl et~al\mbox{.}}{2020}]%
        {GigerlHPMB20}
\bibfield{author}{\bibinfo{person}{Barbara Gigerl}, \bibinfo{person}{Vedad
  Hadzic}, \bibinfo{person}{Robert Primas}, \bibinfo{person}{Stefan Mangard},
  {and} \bibinfo{person}{Roderick Bloem}.} \bibinfo{year}{2020}\natexlab{}.
\newblock \bibinfo{title}{{Coco:} Co-Design and Co-Verification of Masked
  Software Implementations on {CPUs}}.
\newblock , \bibinfo{numpages}{1294}~pages.
\newblock


\bibitem[\protect\citeauthoryear{Goodwill, Jun, Jaffe, and Rohatgi}{Goodwill
  et~al\mbox{.}}{2011}]%
        {GoodwillJJR2011}
\bibfield{author}{\bibinfo{person}{Gilbert Goodwill}, \bibinfo{person}{Benjamin
  Jun}, \bibinfo{person}{Josh Jaffe}, {and} \bibinfo{person}{Pankaj Rohatgi}.}
  \bibinfo{year}{2011}\natexlab{}.
\newblock \showarticletitle{A Testing Methodology for Side-Channel Resistance
  Validation}.
\newblock  (\bibinfo{year}{2011}).
\newblock


\bibitem[\protect\citeauthoryear{Goubin}{Goubin}{2001}]%
        {GoubinCHES2001}
\bibfield{author}{\bibinfo{person}{Louis Goubin}.}
  \bibinfo{year}{2001}\natexlab{}.
\newblock \showarticletitle{A Sound Method for Switching between {Boolean} and
  Arithmetic Masking}. In \bibinfo{booktitle}{\emph{CHES}}.
  \bibinfo{pages}{3--15}.
\newblock
\showISBNx{978-3-540-44709-2}


\bibitem[\protect\citeauthoryear{Hutter and Tunstall}{Hutter and
  Tunstall}{2019}]%
        {HutterJCEN2019}
\bibfield{author}{\bibinfo{person}{Michael Hutter} {and}
  \bibinfo{person}{Michael Tunstall}.} \bibinfo{year}{2019}\natexlab{}.
\newblock \showarticletitle{Constant-time higher-order {Boolean}-to-arithmetic
  masking}.
\newblock \bibinfo{journal}{\emph{Journal of Cryptographic Engineering}}
  \bibinfo{volume}{9} (\bibinfo{date}{06} \bibinfo{year}{2019}).
\newblock
\urldef\tempurl%
\url{https://doi.org/10.1007/s13389-018-0191-z}
\showDOI{\tempurl}


\bibitem[\protect\citeauthoryear{{International Organization for
  Standardization}}{{International Organization for Standardization}}{2016}]%
        {ISO17825}
\bibfield{author}{\bibinfo{person}{{International Organization for
  Standardization}}.} \bibinfo{year}{2016}\natexlab{}.
\newblock \bibinfo{title}{Testing methods for the mitigation of non-invasive
  attack classes against cryptographic modules}.
\newblock \bibinfo{howpublished}{International Standard ISO/IEC 17825:2016(E)}.
\newblock


\bibitem[\protect\citeauthoryear{Ishai, Sahai, and Wagner}{Ishai
  et~al\mbox{.}}{2003}]%
        {IshaiSW03}
\bibfield{author}{\bibinfo{person}{Yuval Ishai}, \bibinfo{person}{Amit Sahai},
  {and} \bibinfo{person}{David~A. Wagner}.} \bibinfo{year}{2003}\natexlab{}.
\newblock \showarticletitle{Private Circuits: Securing Hardware against Probing
  Attacks}. In \bibinfo{booktitle}{\emph{{CRYPTO}}}. \bibinfo{pages}{463--481}.
\newblock


\bibitem[\protect\citeauthoryear{Kocher}{Kocher}{1996}]%
        {Kocher96}
\bibfield{author}{\bibinfo{person}{Paul~C. Kocher}.}
  \bibinfo{year}{1996}\natexlab{}.
\newblock \showarticletitle{Timing Attacks on Implementations of
  {Diffie-Hellman}, {RSA}, {DSS}, and Other Systems}. In
  \bibinfo{booktitle}{\emph{{CRYPTO}}}. \bibinfo{pages}{104--113}.
\newblock


\bibitem[\protect\citeauthoryear{Kocher, Jaffe, and Jun}{Kocher
  et~al\mbox{.}}{1999}]%
        {Kocher1999}
\bibfield{author}{\bibinfo{person}{Paul~C. Kocher}, \bibinfo{person}{Joshua
  Jaffe}, {and} \bibinfo{person}{Benjamin Jun}.}
  \bibinfo{year}{1999}\natexlab{}.
\newblock \showarticletitle{Differential Power Analysis}. In
  \bibinfo{booktitle}{\emph{CRYPTO}}. \bibinfo{pages}{388--397}.
\newblock


\bibitem[\protect\citeauthoryear{Kr{\"a}mer, Nedospasov, Schl{\"o}sser, and
  Seifert}{Kr{\"a}mer et~al\mbox{.}}{2013}]%
        {KN+13}
\bibfield{author}{\bibinfo{person}{Juliane Kr{\"a}mer}, \bibinfo{person}{Dmitry
  Nedospasov}, \bibinfo{person}{Alexander Schl{\"o}sser}, {and}
  \bibinfo{person}{Jean-Pierre Seifert}.} \bibinfo{year}{2013}\natexlab{}.
\newblock \showarticletitle{Differential Photonic Emission Analysis}. In
  \bibinfo{booktitle}{\emph{COSADE}}. \bibinfo{pages}{1--16}.
\newblock


\bibitem[\protect\citeauthoryear{Lai and Massey}{Lai and Massey}{1991}]%
        {Lai90}
\bibfield{author}{\bibinfo{person}{Xuejia Lai} {and} \bibinfo{person}{James~L.
  Massey}.} \bibinfo{year}{1991}\natexlab{}.
\newblock \showarticletitle{A Proposal for a New Block Encryption Standard}. In
  \bibinfo{booktitle}{\emph{Eurocrypt}}. \bibinfo{pages}{389--404}.
\newblock


\bibitem[\protect\citeauthoryear{Lou, Zhang, Jiang, and Zhang}{Lou
  et~al\mbox{.}}{2021}]%
        {abs-2103-14244}
\bibfield{author}{\bibinfo{person}{Xiaoxuan Lou}, \bibinfo{person}{Tianwei
  Zhang}, \bibinfo{person}{Jun Jiang}, {and} \bibinfo{person}{Yinqian Zhang}.}
  \bibinfo{year}{2021}\natexlab{}.
\newblock \showarticletitle{A Survey of Microarchitectural Side-channel
  Vulnerabilities, Attacks and Defenses in Cryptography}.
\newblock \bibinfo{journal}{\emph{CoRR}}  \bibinfo{volume}{abs/2103.14244}
  (\bibinfo{year}{2021}).
\newblock


\bibitem[\protect\citeauthoryear{Mangard, Popp, and Gammel}{Mangard
  et~al\mbox{.}}{2005}]%
        {MangardPG05}
\bibfield{author}{\bibinfo{person}{Stefan Mangard}, \bibinfo{person}{Thomas
  Popp}, {and} \bibinfo{person}{Berndt~M. Gammel}.}
  \bibinfo{year}{2005}\natexlab{}.
\newblock \showarticletitle{Side-Channel Leakage of Masked {CMOS} Gates}. In
  \bibinfo{booktitle}{\emph{{CT-RSA}}}. \bibinfo{pages}{351--365}.
\newblock


\bibitem[\protect\citeauthoryear{McCann, Oswald, and Whitnall}{McCann
  et~al\mbox{.}}{2017}]%
        {McCannOW17}
\bibfield{author}{\bibinfo{person}{David McCann}, \bibinfo{person}{Elisabeth
  Oswald}, {and} \bibinfo{person}{Carolyn Whitnall}.}
  \bibinfo{year}{2017}\natexlab{}.
\newblock \showarticletitle{Towards Practical Tools for Side Channel Aware
  Software Engineering: 'Grey Box' Modelling for Instruction Leakages}. In
  \bibinfo{booktitle}{\emph{{USENIX} Security Symposium}}.
  \bibinfo{pages}{199--216}.
\newblock


\bibitem[\protect\citeauthoryear{Messerges}{Messerges}{2000}]%
        {Mess00}
\bibfield{author}{\bibinfo{person}{Thomas~S. Messerges}.}
  \bibinfo{year}{2000}\natexlab{}.
\newblock \emph{\bibinfo{title}{Power Analysis Attacks and Countermeasures for
  Cryptographic Algorithms}}.
\newblock \bibinfo{thesistype}{Ph.\,D. Dissertation}.
  \bibinfo{school}{{University of Illinois at Chicago, USA}}.
\newblock


\bibitem[\protect\citeauthoryear{Messerges, Dabbish, and Sloan}{Messerges
  et~al\mbox{.}}{1999}]%
        {MessergesDS99}
\bibfield{author}{\bibinfo{person}{Thomas~S. Messerges},
  \bibinfo{person}{Ezzy~A. Dabbish}, {and} \bibinfo{person}{Robert~H. Sloan}.}
  \bibinfo{year}{1999}\natexlab{}.
\newblock \showarticletitle{Power Analysis Attacks of Modular Exponentiation in
  Smartcards}. In \bibinfo{booktitle}{\emph{{CHES}}}.
  \bibinfo{pages}{144--157}.
\newblock


\bibitem[\protect\citeauthoryear{Moradi and Mischke}{Moradi and
  Mischke}{2013}]%
        {MoradiM13}
\bibfield{author}{\bibinfo{person}{Amir Moradi} {and} \bibinfo{person}{Oliver
  Mischke}.} \bibinfo{year}{2013}\natexlab{}.
\newblock \showarticletitle{Comprehensive Evaluation of {AES} Dual Ciphers as a
  Side-Channel Countermeasure}. In \bibinfo{booktitle}{\emph{{ICICS}}}.
  \bibinfo{pages}{245--258}.
\newblock


\bibitem[\protect\citeauthoryear{Moradi, Mischke, and Paar}{Moradi
  et~al\mbox{.}}{2013}]%
        {MoradiMP13}
\bibfield{author}{\bibinfo{person}{Amir Moradi}, \bibinfo{person}{Oliver
  Mischke}, {and} \bibinfo{person}{Christof Paar}.}
  \bibinfo{year}{2013}\natexlab{}.
\newblock \showarticletitle{One Attack to Rule Them All: Collision Timing
  Attack versus 42 {AES} {ASIC} Cores}.
\newblock \bibinfo{journal}{\emph{{IEEE} Trans. Computers}}
  \bibinfo{volume}{62}, \bibinfo{number}{9} (\bibinfo{year}{2013}),
  \bibinfo{pages}{1786--1798}.
\newblock


\bibitem[\protect\citeauthoryear{{National Institute of Standards and
  Technology}}{{National Institute of Standards and Technology}}{2015}]%
        {FIPS1804}
\bibfield{author}{\bibinfo{person}{{National Institute of Standards and
  Technology}}.} \bibinfo{year}{2015}\natexlab{}.
\newblock \bibinfo{booktitle}{\emph{Security Requirements for Cryptographic
  Modules}}.
\newblock \bibinfo{type}{{T}echnical {R}eport} Federal Information Processing
  Standards Publications (FIPS PUBS) FIPS 180-4. \bibinfo{institution}{U.S.
  Department of Commerce}, \bibinfo{address}{Washington, D.C.}
\newblock
\urldef\tempurl%
\url{https://doi.org/10.6028/NIST.FIPS.180-4}
\showDOI{\tempurl}


\bibitem[\protect\citeauthoryear{Nikova, Rechberger, and Rijmen}{Nikova
  et~al\mbox{.}}{2006}]%
        {DBLP:conf/icics/NikovaRR06}
\bibfield{author}{\bibinfo{person}{Svetla Nikova}, \bibinfo{person}{Christian
  Rechberger}, {and} \bibinfo{person}{Vincent Rijmen}.}
  \bibinfo{year}{2006}\natexlab{}.
\newblock \showarticletitle{Threshold Implementations Against Side-Channel
  Attacks and Glitches}. In \bibinfo{booktitle}{\emph{{ICICS}}}.
  \bibinfo{pages}{529--545}.
\newblock


\bibitem[\protect\citeauthoryear{Papagiannopoulos and
  Veshchikov}{Papagiannopoulos and Veshchikov}{2017}]%
        {Papagiannopoulos17}
\bibfield{author}{\bibinfo{person}{Kostas Papagiannopoulos} {and}
  \bibinfo{person}{Nikita Veshchikov}.} \bibinfo{year}{2017}\natexlab{}.
\newblock \showarticletitle{Mind the Gap: Towards Secure 1st-Order Masking in
  Software}. In \bibinfo{booktitle}{\emph{{COSADE}}}.
  \bibinfo{pages}{282--297}.
\newblock


\bibitem[\protect\citeauthoryear{Pardo}{Pardo}{2013}]%
        {PardoScott2013EaNT}
\bibfield{author}{\bibinfo{person}{Scott Pardo}.}
  \bibinfo{year}{2013}\natexlab{}.
\newblock \bibinfo{booktitle}{\emph{Equivalence and Noninferiority Tests for
  Quality, Manufacturing and Test Engineers} (\bibinfo{edition}{1} ed.)}.
\newblock \bibinfo{publisher}{CRC Press LLC}, \bibinfo{address}{Philadelphia,
  PA}.
\newblock
\showISBNx{9781466586888}


\bibitem[\protect\citeauthoryear{Prouff, Rivain, and Bevan}{Prouff
  et~al\mbox{.}}{2009}]%
        {ProuffRB09}
\bibfield{author}{\bibinfo{person}{Emmanuel Prouff}, \bibinfo{person}{Matthieu
  Rivain}, {and} \bibinfo{person}{R{\'{e}}gis Bevan}.}
  \bibinfo{year}{2009}\natexlab{}.
\newblock \showarticletitle{Statistical Analysis of Second Order Differential
  Power Analysis}.
\newblock \bibinfo{journal}{\emph{{IEEE} Trans. Computers}}
  \bibinfo{volume}{58}, \bibinfo{number}{6} (\bibinfo{year}{2009}),
  \bibinfo{pages}{799--811}.
\newblock


\bibitem[\protect\citeauthoryear{Quisquater and Samyde}{Quisquater and
  Samyde}{2001}]%
        {QuisquaterS01}
\bibfield{author}{\bibinfo{person}{Jean{-}Jacques Quisquater} {and}
  \bibinfo{person}{David Samyde}.} \bibinfo{year}{2001}\natexlab{}.
\newblock \showarticletitle{ElectroMagnetic Analysis {(EMA):} Measures and
  Counter-Measures for Smart Cards}. In \bibinfo{booktitle}{\emph{E-smart}}.
  \bibinfo{pages}{200--210}.
\newblock


\bibitem[\protect\citeauthoryear{Renauld, Standaert, Veyrat{-}Charvillon,
  Kamel, and Flandre}{Renauld et~al\mbox{.}}{2011}]%
        {RenauldSVKF11}
\bibfield{author}{\bibinfo{person}{Mathieu Renauld},
  \bibinfo{person}{Fran{\c{c}}ois{-}Xavier Standaert}, \bibinfo{person}{Nicolas
  Veyrat{-}Charvillon}, \bibinfo{person}{Dina Kamel}, {and}
  \bibinfo{person}{Denis Flandre}.} \bibinfo{year}{2011}\natexlab{}.
\newblock \showarticletitle{A Formal Study of Power Variability Issues and
  Side-Channel Attacks for Nanoscale Devices}. In
  \bibinfo{booktitle}{\emph{{EUROCRYPT}}}. \bibinfo{pages}{109--128}.
\newblock


\bibitem[\protect\citeauthoryear{Reparaz}{Reparaz}{2016}]%
        {Reparaz16}
\bibfield{author}{\bibinfo{person}{Oscar Reparaz}.}
  \bibinfo{year}{2016}\natexlab{}.
\newblock \showarticletitle{Detecting Flawed Masking Schemes with Leakage
  Detection Tests}. In \bibinfo{booktitle}{\emph{{FSE}}}.
  \bibinfo{pages}{204--222}.
\newblock


\bibitem[\protect\citeauthoryear{Rivest, Robshaw, Sidney, and Yin}{Rivest
  et~al\mbox{.}}{1998}]%
        {Rivest1998}
\bibfield{author}{\bibinfo{person}{Ronald~L. Rivest}, \bibinfo{person}{M.~J.~B.
  Robshaw}, \bibinfo{person}{R. Sidney}, {and} \bibinfo{person}{Y.~L. Yin}.}
  \bibinfo{year}{1998}\natexlab{}.
\newblock \showarticletitle{The RC6 Block Cipher}. In
  \bibinfo{booktitle}{\emph{in First Advanced Encryption Standard (AES)
  Conference}}. \bibinfo{pages}{16}.
\newblock


\bibitem[\protect\citeauthoryear{Sasdrich, Bock, and Moradi}{Sasdrich
  et~al\mbox{.}}{2018}]%
        {SasdrichB018}
\bibfield{author}{\bibinfo{person}{Pascal Sasdrich},
  \bibinfo{person}{Ren{\'{e}} Bock}, {and} \bibinfo{person}{Amir Moradi}.}
  \bibinfo{year}{2018}\natexlab{}.
\newblock \showarticletitle{Threshold Implementation in Software - Case Study
  of {PRESENT}}. In \bibinfo{booktitle}{\emph{{COSADE}}}.
  \bibinfo{pages}{227--244}.
\newblock


\bibitem[\protect\citeauthoryear{Schneider and Moradi}{Schneider and
  Moradi}{2015}]%
        {SchneiderM15}
\bibfield{author}{\bibinfo{person}{Tobias Schneider} {and}
  \bibinfo{person}{Amir Moradi}.} \bibinfo{year}{2015}\natexlab{}.
\newblock \showarticletitle{Leakage Assessment Methodology - A Clear Roadmap
  for Side-Channel Evaluations}. In \bibinfo{booktitle}{\emph{{CHES}}}.
  \bibinfo{pages}{495--513}.
\newblock


\bibitem[\protect\citeauthoryear{Schuirmann}{Schuirmann}{1987}]%
        {Sch87}
\bibfield{author}{\bibinfo{person}{Donald~J Schuirmann}.}
  \bibinfo{year}{1987}\natexlab{}.
\newblock \showarticletitle{A comparison of the two one-sided tests procedure
  and the power approach for assessing the equivalence of average
  bioavailability}.
\newblock \bibinfo{journal}{\emph{Journal of pharmacokinetics and
  biopharmaceutics}} \bibinfo{volume}{15}, \bibinfo{number}{6}
  (\bibinfo{year}{1987}), \bibinfo{pages}{657--680}.
\newblock


\bibitem[\protect\citeauthoryear{Sehatbakhsh, Yilmaz, Zajic, and
  Prvulovic}{Sehatbakhsh et~al\mbox{.}}{2020}]%
        {SehatbakhshYZP20}
\bibfield{author}{\bibinfo{person}{Nader Sehatbakhsh},
  \bibinfo{person}{Baki~Berkay Yilmaz}, \bibinfo{person}{Alenka~G. Zajic},
  {and} \bibinfo{person}{Milos Prvulovic}.} \bibinfo{year}{2020}\natexlab{}.
\newblock \showarticletitle{{EMSim:} A Microarchitecture-Level Simulation Tool
  for Modeling Electromagnetic Side-Channel Signals}. In
  \bibinfo{booktitle}{\emph{{HPCA}}}. \bibinfo{pages}{71--85}.
\newblock


\bibitem[\protect\citeauthoryear{Shahverdi, Shirinov, and
  Dachman{-}Soled}{Shahverdi et~al\mbox{.}}{2020}]%
        {abs-2006-15007}
\bibfield{author}{\bibinfo{person}{Aria Shahverdi}, \bibinfo{person}{Mahammad
  Shirinov}, {and} \bibinfo{person}{Dana Dachman{-}Soled}.}
  \bibinfo{year}{2020}\natexlab{}.
\newblock \showarticletitle{Database Reconstruction from Noisy Volumes: {A}
  Cache Side-Channel Attack on {SQLite}}.
\newblock \bibinfo{journal}{\emph{CoRR}}  \bibinfo{volume}{abs/2006.15007}
  (\bibinfo{year}{2020}).
\newblock


\bibitem[\protect\citeauthoryear{Shelton, Samwel, Batina, Regazzoni, Wagner,
  and Yarom}{Shelton et~al\mbox{.}}{2021}]%
        {SSB19}
\bibfield{author}{\bibinfo{person}{Madura~A. Shelton}, \bibinfo{person}{Niels
  Samwel}, \bibinfo{person}{Lejla Batina}, \bibinfo{person}{Francesco
  Regazzoni}, \bibinfo{person}{Markus Wagner}, {and} \bibinfo{person}{Yuval
  Yarom}.} \bibinfo{year}{2021}\natexlab{}.
\newblock \showarticletitle{{Rosita:} Towards Automatic Elimination of
  Power-Analysis Leakage in Ciphers}. In \bibinfo{booktitle}{\emph{{NDSS}}}.
\newblock


\bibitem[\protect\citeauthoryear{Shusterman, Kang, Haskal, Meltser, Mittal,
  Oren, and Yarom}{Shusterman et~al\mbox{.}}{2019}]%
        {ShustermanKHMMO19}
\bibfield{author}{\bibinfo{person}{Anatoly Shusterman},
  \bibinfo{person}{Lachlan Kang}, \bibinfo{person}{Yarden Haskal},
  \bibinfo{person}{Yosef Meltser}, \bibinfo{person}{Prateek Mittal},
  \bibinfo{person}{Yossi Oren}, {and} \bibinfo{person}{Yuval Yarom}.}
  \bibinfo{year}{2019}\natexlab{}.
\newblock \showarticletitle{Robust Website Fingerprinting Through the Cache
  Occupancy Channel}. In \bibinfo{booktitle}{\emph{{USENIX} Security
  Symposium}}. \bibinfo{pages}{639--656}.
\newblock


\bibitem[\protect\citeauthoryear{Smith}{Smith}{1997}]%
        {Smith97DSP}
\bibfield{author}{\bibinfo{person}{Steven~W. Smith}.}
  \bibinfo{year}{1997}\natexlab{}.
\newblock \bibinfo{booktitle}{\emph{The Scientist and Engineer's Guide to
  Digital Signal Processing}}.
\newblock \bibinfo{publisher}{California Technical Publishing},
  \bibinfo{address}{USA}.
\newblock
\showISBNx{0966017633}


\bibitem[\protect\citeauthoryear{Standaert}{Standaert}{2018}]%
        {Standaert18}
\bibfield{author}{\bibinfo{person}{Fran{\c{c}}ois{-}Xavier Standaert}.}
  \bibinfo{year}{2018}\natexlab{}.
\newblock \showarticletitle{How (Not) to Use {Welch}'s T-Test in Side-Channel
  Security Evaluations}. In \bibinfo{booktitle}{\emph{{CARDIS}}}.
  \bibinfo{pages}{65--79}.
\newblock


\bibitem[\protect\citeauthoryear{Veshchikov}{Veshchikov}{2014}]%
        {Veshchikov14}
\bibfield{author}{\bibinfo{person}{Nikita Veshchikov}.}
  \bibinfo{year}{2014}\natexlab{}.
\newblock \showarticletitle{{SILK:} high level of abstraction leakage simulator
  for side channel analysis}. In \bibinfo{booktitle}{\emph{PPREW@ACSAC}}.
  \bibinfo{pages}{3:1--3:11}.
\newblock


\bibitem[\protect\citeauthoryear{Welch}{Welch}{1947}]%
        {welch1947generalization}
\bibfield{author}{\bibinfo{person}{Bernard~L Welch}.}
  \bibinfo{year}{1947}\natexlab{}.
\newblock \showarticletitle{The generalization of student's' problem when
  several different population variances are involved}.
\newblock \bibinfo{journal}{\emph{Biometrika}} \bibinfo{volume}{34},
  \bibinfo{number}{1/2} (\bibinfo{year}{1947}), \bibinfo{pages}{28--35}.
\newblock


\bibitem[\protect\citeauthoryear{Yan, Fletcher, and Torrellas}{Yan
  et~al\mbox{.}}{2020}]%
        {YanFT20}
\bibfield{author}{\bibinfo{person}{Mengjia Yan},
  \bibinfo{person}{Christopher~W. Fletcher}, {and} \bibinfo{person}{Josep
  Torrellas}.} \bibinfo{year}{2020}\natexlab{}.
\newblock \showarticletitle{Cache Telepathy: Leveraging Shared Resource Attacks
  to Learn {DNN} Architectures}. In \bibinfo{booktitle}{\emph{{USENIX} Security
  Symposium}}. \bibinfo{pages}{2003--2020}.
\newblock


\bibitem[\protect\citeauthoryear{Zhang, Bao, Lin, Rijmen, Yang, and
  Verbauwhede}{Zhang et~al\mbox{.}}{2015}]%
        {ZhangBLR0V15}
\bibfield{author}{\bibinfo{person}{Wentao Zhang}, \bibinfo{person}{Zhenzhen
  Bao}, \bibinfo{person}{Dongdai Lin}, \bibinfo{person}{Vincent Rijmen},
  \bibinfo{person}{Bohan Yang}, {and} \bibinfo{person}{Ingrid Verbauwhede}.}
  \bibinfo{year}{2015}\natexlab{}.
\newblock \showarticletitle{{RECTANGLE:} a bit-slice lightweight block cipher
  suitable for multiple platforms}.
\newblock \bibinfo{journal}{\emph{Sci. China Inf. Sci.}} \bibinfo{volume}{58},
  \bibinfo{number}{12} (\bibinfo{year}{2015}), \bibinfo{pages}{1--15}.
\newblock


\end{thebibliography}

\begin{appendices}
\crefalias{section}{appendix}

\end{appendices}
\end{document}